\documentclass[preprint, onecolumn, showpacs, showkeys]{revtex4-1}
\usepackage{amsmath}
\usepackage{amssymb}
\usepackage{graphicx}
\usepackage{hyperref}
\usepackage{subfigure}
\usepackage{setspace}
\usepackage{color}

\newcommand{\mean}[1]{\langle #1 \rangle}

\newcommand{\vel}{\mathrm{v}}
\newcommand{\bx}{\boldsymbol{x}}
\newcommand{\bq}{\boldsymbol{q}}
\newcommand{\bnd}[1]{\tilde{\boldsymbol{#1}}}
\newcommand{\txpar}{\tilde x_\parallel}
\newcommand{\tqpar}{\tilde q_\parallel}
\newcommand{\txper}{\tilde x_\perp}
\newcommand{\tqper}{\tilde q_\perp}
\newcommand{\tbxper}{\tilde{\boldsymbol{x}}_\perp}
\newcommand{\tbqper}{\tilde{\boldsymbol{q}}_\perp}
\newcommand{\tbx}{\tilde{\boldsymbol{x}}}
\newcommand{\tbq}{\tilde{\boldsymbol{q}}}
\newcommand{\tx}{\tilde x}
\newcommand{\tq}{\tilde q}
\newcommand{\tv}{\tilde \vel}

\begin{document}

\title{ Striated Populations in Disordered Environments with Advection}
\author{Thiparat Chotibut}
\email{Electronic address: thiparatc@gmail.com}
\author{David R. Nelson}
\email{Electronic address: nelson@physics.harvard.edu}
\affiliation{Department of Physics, Harvard University, Cambridge, Massachusetts
02138, USA}
\author{Sauro Succi}
\email{Electronic address: succi@iac.cnr.it}
\affiliation{Istituto per le Applicazioni del Calcolo, Rome Via dei Taurini
19, 00185, Roma, Italy,}
\affiliation{Institute for Applied Computational Science, John A. Paulson
School of Engineering and Applied Science, Harvard University, Cambridge,
Massachusetts 02138, USA}
\date{\today}
\begin{abstract}
Growth in     static and  controlled environments such as a Petri dish can
be used to study the spatial population dynamics of microorganisms. However,
natural populations such as marine microbes experience   fluid advection
and often grow up in heterogeneous environments.
We investigate a generalized Fisher-Kolmogorov-Petrovsky-Piscounov  (FKPP)
equation describing
single species population
subject to a constant flow field and  quenched random
spatially inhomogeneous growth rates with a fertile overall growth condition.
  We  analytically and numerically demonstrate
  that the  non-equilibrium steady-state population density  develops
a flow-driven striation pattern. The striations are highly asymmetric
with a longitudinal correlation length that
diverges linearly with the flow speed and a transverse correlation  length
that approaches a finite velocity-independent value. Linear response
theory is developed to study  the statistics of the steady states. Theoretical
predictions  show
excellent agreement with the numerical steady
states of the   generalized FKPP equation obtained from  Lattice
Boltzmann simulations.
These findings suggest that, although the growth disorder can be spatially
uncorrelated, correlated population structures with striations emerge naturally at sufficiently
strong advection.
\end{abstract}
\maketitle
\tableofcontents
\section{Introduction}
\label{sec:intro}

How growth, competition,  and   dispersal affect the spatial structure of
living populations
has been  a keystone to understanding biodiversity and stability of ecosystems
\cite{May:1990aa,Gotelli:2001aa,Murray:2001aa,Rietkerk:2004aa,Rietkerk:2008aa}.
Numerous works demonstrate that simple  dispersal mechanisms, such as diffusive
spreading and chemotaxis, when coupled with spatial heterogeneity of resources,
may result in intricate far-from-equilibrium  spatial patterning in diverse
populations, including, for example, bacterial patterns on a Petri dish \cite{Budrene:1991aa,Brenner:1998aa,Cates:2010aa},
soft-sediment  mussel  beds \cite{Liu:2012aa,Liu:2014aa,Koppel:2005aa}, and
vegetation patterns in arid ecosystems \cite{Von-Hardenberg:2001aa,Rietkerk:2004aa,Borgogno:2009aa}.
More complicated dispersal mechanisms such as advective transport may also
 influence  spatial distributions in natural populations.
For instance, patchiness and filamentation of planktonic communities on ocean
surfaces can arise from the delicate interplay between growth, competition,
diffusion,
and advection by turbulence
and chaotic oceanic flows \cite{Tel:2005ab,Neufeld:2012aa,McKiver:2009aa,Liu:2014aa}.
 Two-dimensional compressible turbulence restricts growth and competition
to  thin filaments leading to a dramatic decrease in the global carrying
capacity
in model planktonic populations \cite{Pigolotti:2012aa,Benzi:2012aa}.   Even
in
a simple effectively unidirectional flow field,  the combined effect of spatial
variations in the resource and advection on the spatial structure of populations
 is very rich \cite{Nelson:2012aa,Dahmen:1999aa,Dahmen:2000aa,Desai:2005aa,Nelson:1998aa,Lin:2004aa,Lutscher:2010aa,Pachepsky:2005aa,Pringle:2011aa,Lutscher:2007aa,Kogan:2015aa}.

A general  framework for  systematic exploration
of the growth
and diffusion of single-species populations in an advective, spatially-varying growth
environments is the reaction-diffusion-advection
 equation \cite{Neufeld:2009aa,Murray:2001aa}:
\begin{equation}
\label{eqn: More_Gen_FKPP}
\partial_t c+ \nabla \cdot \left[ \boldsymbol\vel(\boldsymbol x,t)  c \right]
= \nabla \cdot \left[ D(\boldsymbol
x,t) \nabla c \right ] + f[c].
\end{equation}
 In an ecological context, Eq. (\ref{eqn: More_Gen_FKPP}) prescribes the
time evolution of the coarse-grained
density $c\equiv c(\boldsymbol x,t)$ of populations consisting of individuals
that are
transported by  advection with the velocity field $\boldsymbol\vel(\boldsymbol
x,t),$
diffuses (for example by, say, a run and tumble mechanism in the case of bacteria)
with a space and time-dependent diffusivity $D(\boldsymbol
 x,t),$ and  reproduce with the density-dependent growth
rate $f[c ]$ that depends on a local environment.
When advection is absent, the diffusivity
is constant
  $D(\boldsymbol x,t) = D$, and the resources necessary for
growth are homogeneously distributed, the density of populations that reproduce
with the logistic growth rate
$f[c] = ac-bc^2$  (with $a$ and $b$ constant) obey the well-known Fisher-Kolmogorov-Petrovsky-Piscounov
 (FKPP) equation:
\begin{equation}
\label{eqn: FKPP}
\partial_t c  = D \nabla^2 c + ac-bc^{2},
\end{equation}
which, for a  uniform positive growth rate $a$,  admits a stable non-linear
wave solution describing  a spatial range expansion of populations that advance
from a  region  saturated at the local carrying
capacity $K \equiv a/b,$ into  unoccupied territory \cite{Fisher:1937aa,KPP:1937,Murray:2001aa,Murray:2002aa}.
At long times, reproductive growth and diffusion fill up and saturate the
domain; the
population density becomes  featureless and is equal to the
 carrying capacity $K,$ which is the stable steady state of   Eq. (\ref{eqn:
FKPP}). 

In contrast, interesting phenomena and non-trivial spatial population structure
at long times arise
when a constant flow field  $\boldsymbol\vel(\boldsymbol x,t) = \boldsymbol \vel$ advects the population
across a spatially quenched growth landscape of the form $f[c]= a(\boldsymbol
x)c -bc^{2}$
\cite{Nelson:2012aa,Dahmen:1999aa,Dahmen:2000aa,Nelson:1998aa}. This minimal
extension of the FKPP equation reads 
\begin{equation}
\label{eqn: Gen_FKPP}
\partial_t c + \boldsymbol\vel\cdot \nabla c = D \nabla^2 c + a(\boldsymbol
x)c-bc^{2}.
\end{equation}
Since advection  can wash away the populations from the favourable  growth
hot spots and diffusion tends to spread populations out, the  density $c(\boldsymbol
x,t)$ deviates from the profile
of the  local carrying capacity
$K(\boldsymbol x) \equiv a(\boldsymbol x)/b$ in regions where $a(\boldsymbol x) > 0.$   
 For instance, in a landscape with a  finite-size favourable growth hot spot
 surrounded  by an unfavourable growth environment, the long-time population
structure that would otherwise localize near the hot spot, can be driven
by advection  to delocalize or becomes extinct \cite{Dahmen:2000aa}. This
localization-extinction transition has been  observed  in microbial
experiments
\cite{Lin:2004aa, Neicu:2000aa}. In addition, in a random environment where
the local growth
rate can be regarded
as a spatially quenched time-independent random variable, early-time growth
eigenfunctions
of the linearized growth operator of Eq. (\ref{eqn: Gen_FKPP}) can exhibit
a localization-delocalization transition, with a remarkable associated non-Hermitian
spectral property \cite{Nelson:1998aa,Molinari:2009aa,Hatano:1998aa,Hatano:1996aa,Shnerb:1998aa,Lutscher:2007aa}.
In  strongly advective environments in which the spatially averaged growth
rate is positive,
all the growth eigenfunctions are delocalized and the spreading dynamics
of populations at the  frontier in the direction transverse to the flow is expected to be super-diffusive  \cite{Nelson:1998aa}. 

While
the early-time properties of  growth eigenfunctions and growth dynamics in
a spatially quenched random growth rate are relatively well understood, less
is known about the structure of long-time steady-state
populations. Close to extinction ( in an overall \textit{hostile} growth
environment in which the growth rates $a(\boldsymbol x)$ are mostly negative), when the
early-time growth eigenfunctions do not  significantly overlap,  the
population
structure at long times  is  tracked by the early-time growth eigenfunctions.
However, more generally,
 significant changes due to  the mode-coupling triggered by the non-linear
saturation term $-bc^2$
 distort this picture, especially when advection is strong \cite{Nelson:1998aa}.

In this work,  with the goal of determining  how  constant advection combines
with
spatial variations in an overall \textit{fertile}  growth landscape to affect
the
long-time population structure,  we  analytically
and numerically study 
steady-state populations described by  Eq. (\ref{eqn: Gen_FKPP}).  Specifically,
we study the  long-time
steady-state
population density $c^*(\boldsymbol
x) \equiv c(\boldsymbol x,t\rightarrow \infty)$ satisfying the nonlinear equation,  
\begin{equation}
\label{eqn: SteadyState}
0=D \nabla^2 c^{*}(\boldsymbol x)-\boldsymbol\vel\cdot \nabla c^{*}(\boldsymbol
x) +
a(\boldsymbol x)c^{*}(\boldsymbol x)-bc^{*2}(\boldsymbol x),
\end{equation}arising from  a weak spatially-quenched random growth landscape
of the form
\begin{equation}
a(\boldsymbol x) \equiv \ a_0 +  \delta a(\boldsymbol x), 
\end{equation} 
where $a_0$ is a constant \textit{positive}  background
growth rate, and  $\delta a(\boldsymbol x)$ is a weak  frozen-in spatial
perturbation such that $\mean{a(\boldsymbol x)} > 0$, where $\mean{\cdot}$
represents a spatial average. In the calculations that follow, we shall draw
$\delta a(\boldsymbol x)$ from a uniform, symmetrical box distribution, 
 $\delta
a(\boldsymbol x) \in \left[ -\Delta,\Delta\right].$ Weak disorder then means
$\left|  \delta
a(\boldsymbol x) \right| /a_0 \ll 1.$ Although
recent investigations  suggest that demographic fluctuations due to stochasticity
in the discrete birth and death events can be important  close
to extinction or in diluted populations \cite{Kessler:2009aa,Juhasz:2013aa,Geyrhofer:2013aa,Geyrhofer:2013aa,Missel:2009aa},
we focus here on the regime where the local carrying capacity $K(\bx)=a(\bx)/b
$ is large, so that demographic fluctuations 
 are small and the mean field
description of the steady state embodied in Eq.(\ref{eqn: SteadyState}) is
reliable.

In  Sec. \ref{sec:
linear_response}. we show that the steady-state density fluctuations from
the mean carrying capacity $\bar K \equiv a_0/ b$ can be regarded as a
linear response to a small  perturbation caused by the quenched random background
growth rate. 
The Green's function
of the associated reaction-diffusion-advection operator,  which we calculate
in $d$ dimensions,
describes the steady-state density response  to a point-like growth
hot spot. In the absence of advection,  the response decays isotropically,
with the diffusive localization length
 $\xi_D \equiv \sqrt{D/a_0},$ from the center of a growth hot spot.  We
then discuss how  advection breaks isotropy:  advection elongates (shortens) the downstream (upstream) longitudinal
localization length, while  symmetrically contracting  the transverse localization length.
Figs. \ref{fig: gf1d} and
\ref{fig: gf2d} illustrate this effect, in $d=1$ and $d=2$, respectively.
   For strong advection such that $\tilde \vel \equiv \vel/\vel_{F} \gg 1$
where $\vel_F
\equiv 2\sqrt{D a_0}$ is the characteristic Fisher speed associated with
the average growth rate \cite{Murray:2001aa,Murray:2002aa},
the downstream and upstream longitudinal localization lengths scale as  $\xi_\parallel^+
\approx \tilde \vel\xi_D = \vel/2a_0$ and  $\xi_\parallel^-
\approx (1/  \tilde \vel)\xi_D = 2D/ \vel$ , while the transverse localization length
scales as $\xi_\perp \approx (1/\tilde \vel) \xi_D.$
The asymmetric elongation and contraction occur simultaneously with the
decay
in the response amplitude, a consequence of the conservation  of
density fluctuations   discussed at the end of Sec. \ref{sec:
linear_response}. 
\begin{figure*}[ht!]
\includegraphics[width = \linewidth]{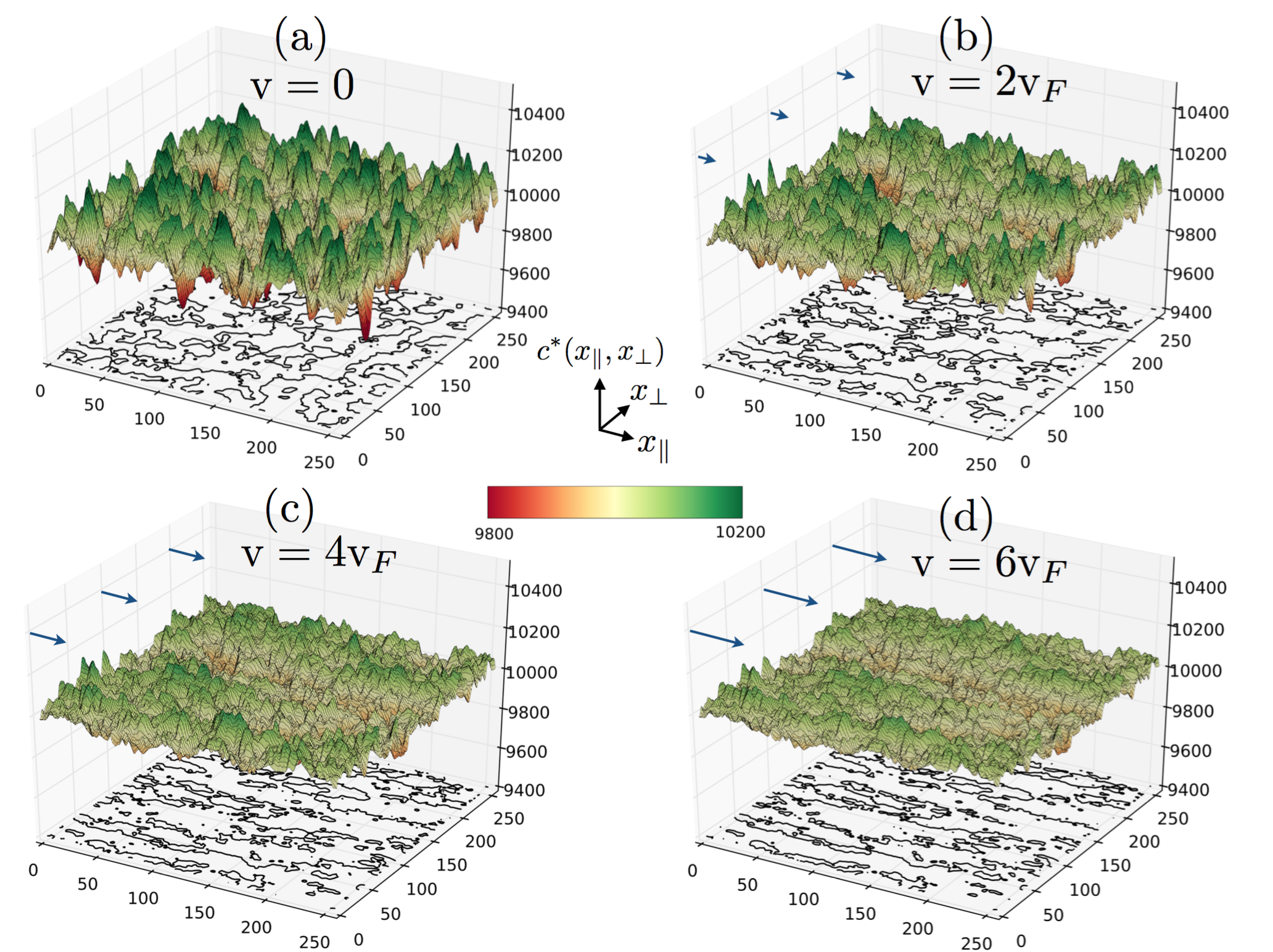}\caption{
(Color online) Typical long-time steady states $c^*(\bx)$ with striations
in two dimensions
that arise from the generalized FKPP equation with periodic boundary conditions
and variable strength advection in the $+\hat
x_\parallel$ direction and with the same realization of a weak spatially
quenched random growth
rate. The black lines at the bottom indicate contours  of $c^*(\bx) $ at
the value of the mean carrying capacity  $\bar K = 10^4$, separating the
more favourable growth domains from the less favourable ones. These plots
result from the lattice Boltzmann simulation discussed in Sec. \ref{sec:
numerical_sim}. Without advection,  as in (a), steady state density exhibit
mild
fluctuations around the mean carrying capacity  $\bar K = 10^4.$
 The correlations of the density fluctuations without advection are isotropic.
However, advection breaks statistical isotropy of the steady state;   correlation
length is
 developed and is elongated in the longitudinal direction, as in (b),
(c) and (d).  Striated patterns  emerge at strong flow
$ \vel \gg \vel_F$ with the longitudinal correlation
length of order the system size while the transverse correlation length remains
finite,  see (d). Note also that strong advection lowers the amplitude of
the density
fluctuations.}
\label{fig: 2dsteadystates}   
\end{figure*}

 In Sec. \ref{sec: striation}, we show  that advection and spatially
quenched \textit{uncorrelated} random growth rates
together  lead to  \textit{striated} patterns  of steady-state population
density in two dimensions, as depicted
 in Fig. \ref{fig: 2dsteadystates}.   Although
the growth rate fluctuations are spatially uncorrelated and the transverse
localization length of the response
from an isolated  growth hot spot shrinks  to zero as advection
becomes stronger,  a finite  transverse correlation
length of the steady state density nevertheless emerges at strong advection.
 Theoretical analysis
 of the two-point
correlation function leading to  striated population structures is
provided in Sec. \ref{sec: two-point
corfunc}. There, we also show that, for  strong advection $ \vel \gg \vel_{F}$,
the longitudinal
correlation length elongates without bound as $\xi_\parallel = \vel/a_0$
whereas
the transverse correlation length approaches the limiting value
$\xi_\perp=\xi_D=\sqrt{D/a_0}$. Consequently, the population structure
becomes highly anisotropic, with the ratio of
 correlation lengths  given by $\xi_\parallel/\xi_\perp = 2 \tilde \vel.$
  Sec. \ref{sec: numerical_sim} compares 
the long distance correlations  from  theoretical analysis with  those from
the lattice Boltzmann simulations. Concluding remarks appear in Sec. \ref{sec:
conclusion}, and detailed calculations of Green's functions, correlations
and structure functions are contained
  in Appendix A.

\section{Steady-state density from the linear response approximation}
\label{sec: linear_response}
To determine how weak spatial
perturbations in the growth rate alter a steady-state density
that is
otherwise  homogeneous and equal to the mean carrying capacity $\bar K= a_0/
b$, we introduce the   density deviation from the steady state, normalized
by
the mean carrying capacity, as follows:
\begin{equation}\label{eqn: def_densfluc}
\phi(\boldsymbol x) \equiv \frac{c^*(\boldsymbol x) - \bar K}{\bar K},
\end{equation}
henceforth referred to as a (static) density fluctuation. Upon substituting
the steady-state density of Eq.
(\ref{eqn: def_densfluc}) into Eq. (\ref{eqn: SteadyState}), we find that
this (static) density fluctuation satisfies 
\begin{equation}\label{eqn: densfluc_allorder}
\left(-D\nabla^2 +\boldsymbol\vel \cdot \nabla+ a_0 \right)\phi(\boldsymbol{x})=
\delta a(\boldsymbol
x)+[\delta a(\boldsymbol x)\phi (\boldsymbol{x})-a_0 \phi^{2}(\boldsymbol{x})].
\end{equation}
 We now establish the linear response theory for  the case of weak spatial
fluctuations in the growth rate. First, we define  $\varepsilon \equiv (\Delta/a_0)
\ge |\delta a(\boldsymbol
x)/a_{0}|$, a dimensionless measure of growth rate fluctuations. 
Observe that, in the absence of both advection
and diffusion, non-zero populations at any point   $\boldsymbol
x$ will grow   and saturate   according to the
local logistic growth process; the steady state density is then given by
the local
 carrying capacity: \begin{equation}\label{eqn: local_car_cap}
c^*(\boldsymbol x)=K(\boldsymbol
x) \equiv  a(\boldsymbol x)/b.
\end{equation}  
Eq. (\ref{eqn:
def_densfluc}), Eq. (\ref{eqn: local_car_cap}), and the bound associated
with uniform distribution $\delta a(\boldsymbol x) \in \left[-\Delta,\Delta\right]$,
implies that the density fluctuations obey $|\phi(\boldsymbol
x)| \lesssim \varepsilon.$  In the presence of either diffusion
or advection,  populations traverse  longer distances and sample a spatial
average of local growth
rates;  the steady state density is then smoothed out, and the
condition $|\phi(\boldsymbol x)| \lesssim\ \varepsilon$   should remain approximately
valid.
In fact,
when diffusion or advection become
strong,    numerical simulations reveal the suppression of density fluctuations,
as shown for strong advection in
 Figs. \ref{fig: 2dsteadystates}
and \ref{fig: 1D_xi_pm} in two dimensions and one dimension, respectively
(the case
of suppression by strong diffusion is similar.) Upon dividing Eq. (\ref{eqn:
densfluc_allorder})
by $a_0$ and noting that  $|\phi(\boldsymbol x)| = O(\varepsilon),$
 we see that the terms in the square bracket scale as $\varepsilon^2$,
 whereas the
other terms scale as $\varepsilon.$ Hence, in the limit of small   $\varepsilon
$, we can  linearize Eq. (\ref{eqn: densfluc_allorder}):
\begin{equation}\label{eqn: ss_linear_response}
\left(-D\nabla^2 +{\bf v} \cdot \nabla+ a_0 \right)\phi(\boldsymbol{x})
= \delta a(\boldsymbol{x}).
\end{equation}
Thus, in this linear approximation (used throughout this paper), static density fluctuations are generated in response to the growth
disorder $\delta a (\boldsymbol x)$ acting as a source term. As
we  show later, Eq. (\ref{eqn: ss_linear_response}) becomes a better and better approximation upon increasing either the $D$
or $|\bf v|.$  \newline
\indent It is  convenient  to  introduce
the non-dimensionalized linear response
equation
 to simplify further calculations.
Upon defining the diffusion length in a growth time
\begin{equation}
\xi_{D} \equiv \sqrt{D/a_0},
\end{equation}
and rescaled quantities
\begin{align} 
\tilde \bx &\equiv \bx/\xi_D, \label{eqn: non-dim-x} \\
\tilde \vel &\equiv \vel/2\sqrt{Da_0} = \vel/\vel_{F},\\
U(\boldsymbol x) &\equiv\delta a(\boldsymbol x)/a_0,
\end{align}
Eq. (\ref{eqn: ss_linear_response}) takes the dimensionless form\begin{equation}\label{eqn:
ss_linear_response_non-dim}
\left(-\tilde \nabla^2 +2\bnd{\vel} \cdot  \tilde \nabla+1 \right)\phi( 
\bnd{x}) = U( \boldsymbol{\tilde x}),
\end{equation}
where $\tilde\nabla$ denotes a gradient with respect to 
 $\bnd{x} $.  The density fluctuations are then given by the
 convolution \begin{equation}\label{eqn:
superposition}
\phi( \boldsymbol{\tilde x}) = \int  G(\boldsymbol{\tilde x}- \bnd{y})U(\bnd{y})
\ d^d \bnd{y},
\end{equation}
where $G( \boldsymbol{ \tilde x} - \bnd{y})$ is a reaction-diffusion-advection
Green's
function that 
satisfies \newline $\left(-\tilde \nabla^2 +2\bnd{\vel} \cdot  \tilde \nabla+1 \right)G(
\bnd{x} - \bnd{y}) \propto  \delta^d( \boldsymbol{\tilde x }- \boldsymbol{\tilde
y }).$
The steady-state population density, corrected for diffusion and advection,
then reads 
\begin{equation}\label{eqn: steadystatefrom_densfluc}
c^*(\tbx ) = \bar K\left[1+ \int  G(\boldsymbol{\tilde x}- \bnd{y})U(\bnd{y})
\ d^d \bnd{y} \right].
\end{equation}
And we must now determine $G(
\bnd{x} - \bnd{y})$.
\subsection{Density modulation due to a point-like growth hot
spot}\label{sec: GF}

 The Green's function  of the reaction-diffusion-advection operator of Eq.
(\ref{eqn: ss_linear_response_non-dim}) describes the response to a Dirac
delta function source term,  a point-like growth hot spot.
In our rescaled coordinates, the Green's
function  satisfies 
\begin{equation}\label{eqn: GF_EqnFor_non-d}
\left(-\tilde \nabla^2 +2\bnd{\vel} \cdot  \tilde \nabla+1 \right)G(
\bnd{x}) = \frac{\Delta}{a_0} \delta ^d (\bnd{x}),
\end{equation}which can be simplified via the substitution
\begin{equation}\label{eqn: gauge_trans}
f_{\tilde \vel}(\bnd{x})\equiv\exp(-{\tilde \vel \tilde x_\parallel})f(\bnd{x}),
\end{equation}
where $x_\parallel$ is the direction along the advective flow. The result
is a  Helmholtz equation with the minus sign in the
Laplacian:
\begin{equation}\label{eqn: GF_EqnFor_gauged}
\left[-\tilde \nabla^2 +(1+\tilde \vel^2) \right] G_{\tilde\vel}(
\bnd{x}) = \frac{\Delta}{a_0} \delta ^d (\bnd{x}).
\end{equation}
In $d$ dimensions, the
isotropic solution of Eq. (\ref{eqn: GF_EqnFor_gauged}), such that $G_{\tilde\vel}(\tilde
\bx)$ vanishes as $\tx \rightarrow \infty$, reads
\cite{Hassani:2013aa}
\begin{align}\label{eqn: GF_dimensionless_gauged}
G_{\tilde\vel}(\tilde \bx) = &\frac{\Delta}{a_0}\left(\frac{1}{\pi}\right)^{d/2}
\nonumber\\
&\times\left[\left(\frac{|\tilde \bx|}{\sqrt{1+\tilde \vel^2}}\right)^{1-d/2}
K_{1-d/2}\left(\sqrt{1+\tilde \vel^2}|\bnd{x}|\right)\right],
\end{align}
 so that\begin{align}
G(\tilde \bx) = &\exp(\tilde \vel \tilde x_\parallel)G_{\tilde\vel}(\bnd{x}).\label{eqn:
GF_gauged_relation}
\end{align}
 Eqs. (\ref{eqn: GF_dimensionless_gauged}) and (\ref{eqn: GF_gauged_relation})
are derived via direct Fourier transformation in Appendix \ref{sec: detailed_calculation}.

Eq. (\ref{eqn: GF_gauged_relation}) encapsulates how advection breaks isotropy:
by enhancing downstream response and suppressing upstream response in Eq. (\ref{eqn: GF_dimensionless_gauged}). 
Since   
$K_{\alpha}(x)=\sqrt{\frac{\pi}{2x}}\exp(-x)\left[ 1 + \frac{4\alpha^2-1}{8x}
+\mathcal{O}(x^{-2})\right]$ for large $x$ \cite{Abramowitz:1964aa}, Eq.(\ref{eqn: GF_gauged_relation})
implies that  $G(\bnd{x})$ is exponentially  localized in all 
transverse directions with the  transverse localization length $\tilde
\xi_{\perp}
=1/\sqrt{1+\tilde
\vel^2}$ 
similar to the localization length of    $G_{\tilde
\vel}(\bnd{x})$, i.e., 
\begin{equation}
G(\tilde x_\parallel=0, \bnd{x}_\perp) \sim \exp \left(-\sqrt{1+\tilde \vel^2}|\bnd{x}_\perp| \right).
\end{equation}
For  $\tilde \vel \gg 1, $ $\tilde \xi_{\perp}$ contracts as $1/\tilde \vel.$
However,  the longitudinal localization lengths are asymmetric with an elongation
in the downstream direction and a contraction in the upstream direction.
 For $\tilde \vel \gg 1,$   
\begin{align}\label{eqn: GF_longitudinal}
G(\tilde x_\parallel, \bnd{x}_\perp=\boldsymbol 0) &\sim \exp(\tilde \vel
\tilde x_\parallel)\exp(-\sqrt{1+\tilde \vel^2}|\tilde x_\parallel|) \nonumber
\\
&=\exp \left[(\tilde x_\parallel - |\tilde x_\parallel|) \tilde \vel - \frac{|\tilde
x_\parallel|}{2 \tilde \vel} + O\left(\frac{1}{\tilde \vel^3} \right) \right];
\end{align}
thus, the downstream ($\txpar > 0 $) and the upstream ($\txpar < 0 $) localization
lengths are given by $\tilde
\xi^+_\parallel \approx 2\tilde \vel$ and $\tilde \xi^-_\parallel
\approx 1/2\tilde \vel,$ respectively.
This steady state asymmetry, with
a contraction in the transverse direction, differs from the early
time growth dynamics, where diffusion  spreads out the population superdiffusively in the
transverse direction \cite{Dahmen:1999aa,Nelson:2012aa,Nelson:1998aa}.

\begin{figure}[!ht]
\includegraphics[width =0.8\linewidth]{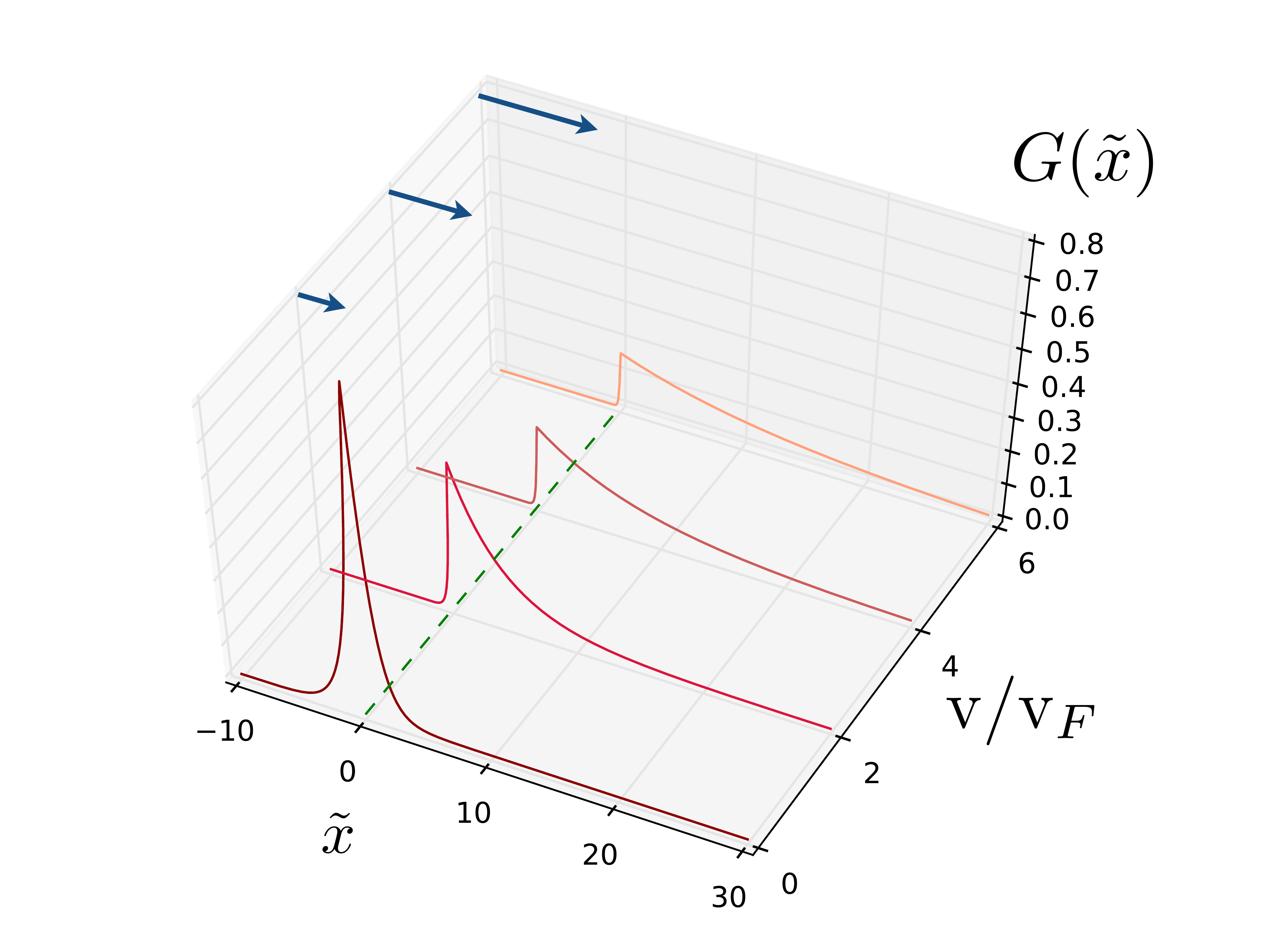}
\caption{(Color online) One-dimensional Green's functions given by Eq.(\ref{eqn: GF_1d})
which describe
the  density fluctuations as a  response to a point-like growth hot spot,
introduced at the origin and indicated as the dashed green line. The response
becomes strongly asymmetric as $\tilde \vel$ increases, with the  downstream
localization length and the upstream localization length that scale respectively
as $\tilde
\xi^+_\parallel = 2\tilde \vel$ and $\tilde
\xi^-_\parallel = 1/2\tilde \vel$ for $\tilde \vel \gg 1.$ Note that the
response amplitude
decays to compensate for the elongated downstream propagation. In this plot,
the strength of
the hot spot $\Delta /a_0$ is set to unity. } 
\label{fig: gf1d}         
\end{figure}

\begin{figure}[!h]     
\includegraphics[width =\linewidth]{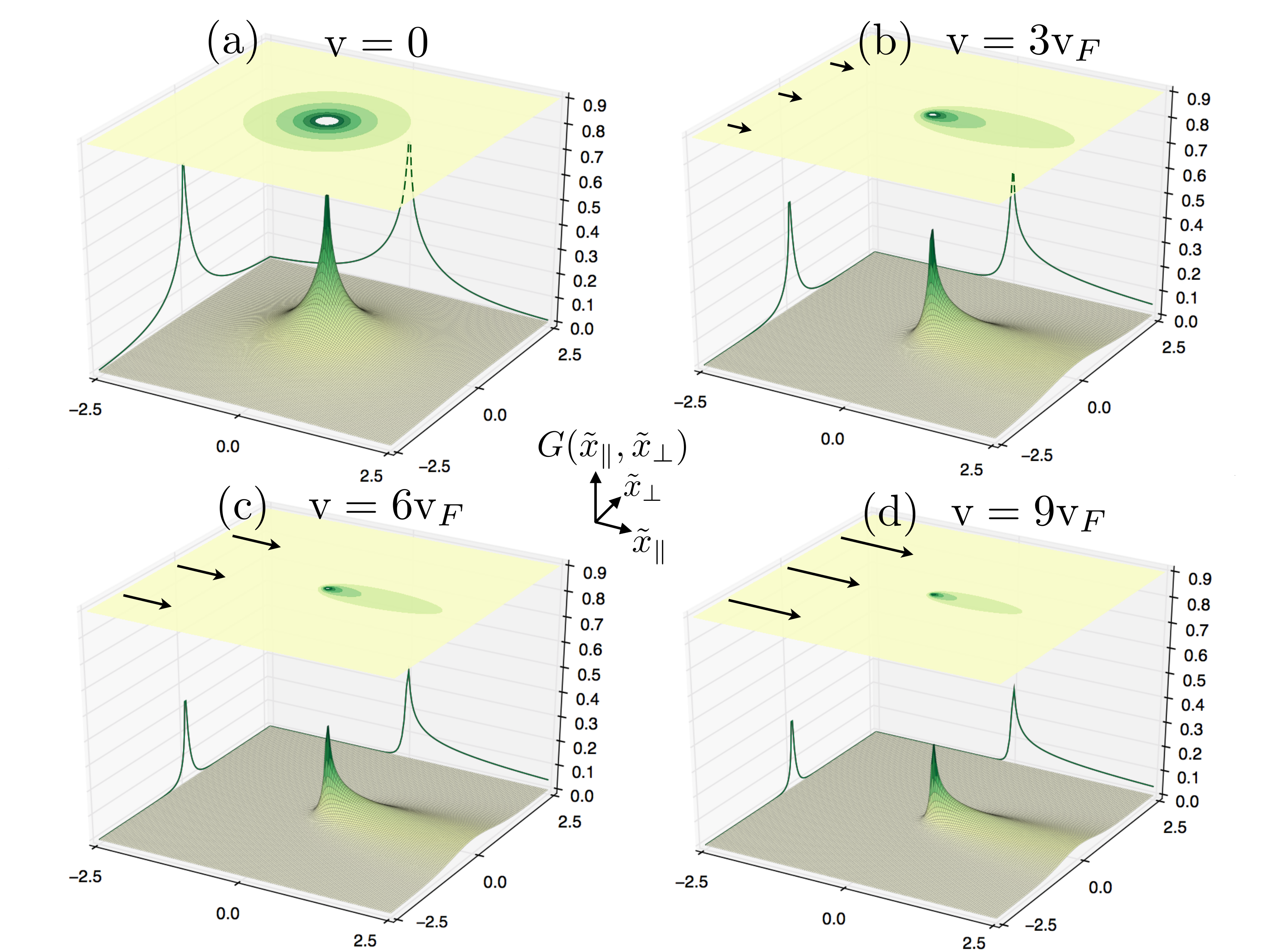}\caption{
(Color online) Two-dimensional Green's function   of  Eq.(\ref{eqn:
2dGF}). 
The response to a point-like growth hot spot is isotropic in the absence
of flow, as represented in
(a), but streaked out in the direction parallel to the flow with an elongated
downstream localization length, as shown in (b)-(d). With increasing flow
speed,
 the overall response amplitude decays. In addition, the transverse response,
controlled by $G_{\tilde
\vel}(\bnd{x}) \sim K_{0}\left(\sqrt{1+\tilde
\vel^2}|\tilde {\bx}|\right)$, contracts
symmetrically,  decaying exponentially with the localization
length
$\xi = 1/\sqrt{1+\tilde
\vel^2}$.   Similar to one dimension,  $\tilde
\xi^+_\parallel = 2\tilde \vel$ and $\tilde
\xi^-_\parallel = 1/2\tilde \vel$ in the flow direction for $\tilde \vel
\gg 1.$ Solid
lines
shown on the plane  $\tilde x_{\parallel} = -2.5 $ and $\tilde x_\perp =2.5$
are  cross-sections of the response along the plane  $\tilde x_\parallel
= 0$ and the plane   $\tilde
x_\perp = 0$, respectively. These cross-sections reveal the behavior of localization
lengths at different velocities. The boundaries of the colored  contour
above each plot correspond to the  Green's function contours  $0.0,0.05,0.1,0.15,0.2,0.25$
and $0.3$, with the color  ranging from yellow to dark green, respectively.
The reduction in
the area of these colored contours at stronger advection reflects the
decay of the response amplitude. In this plot,
the amplitude of
the $\delta$-function hot spot $\Delta /a_0$ is set to 1. }   
      \label{fig: gf2d}
\end{figure}

The steady-state response to advection is, however, constrained
by a  conservation law.  By integrating  Eq. (\ref{eqn: GF_EqnFor_non-d})
over the whole
domain with a periodic boundary
condition,
one finds that \begin{equation}\label{eqn: cons_law}
\int_\Omega G(\bnd{x}) d^d\bnd{x} = \frac{\Delta}{a_0}.
\end{equation}
As a result, the overall response is suppressed  at strong advection: the
response amplitude decreases to compensate for elongated
downstream amplitude.    Suppression of the response amplitude as a tradeoff
for elongated downstream propagation is readily verified in one-dimension.
In this case, the  modified Bessel function takes the simple
form $K_{1/2}(x)=\sqrt{\frac{\pi}{2x}}\exp(-x)$ \cite{Abramowitz:1964aa},
and Eqs. (\ref{eqn: GF_dimensionless_gauged})-(\ref{eqn: GF_gauged_relation})
give \begin{align}
\label{eqn: GF_1d}
G(\tilde x)=\frac{\Delta }{a_0}\frac{1}{\sqrt{2}}\frac{\exp(\tilde
\vel
\tilde x-\sqrt{1+\tilde
\vel^2}|\tilde x|)}{\sqrt{1+\tilde \vel^2}}.
\end{align} 
Hence, advection leads to anisotropic response whose amplitude at  $\tilde
x = 0$ 
decays as $1/\sqrt{1+\tilde \vel^2}$. Fig. \ref{fig: gf1d}. shows $G(\tilde x)$ in one dimension for a variety of velocities. In two dimensions, Eqs. (\ref{eqn: GF_dimensionless_gauged})-(\ref{eqn:
GF_gauged_relation})
give
\begin{equation}\label{eqn: 2dGF}
G(\tilde x_\parallel, \tilde x_\perp) = \frac{\Delta}{a_0}\left(\frac{1}{\pi}\right)
\exp(\tilde \vel \tilde x_\parallel)K_{0}\left(\sqrt{1+\tilde \vel^2}|\tilde {\bx}|\right).
\end{equation}
Although there is a logarithmic singularity at $|\tilde {\bx}|=0$ (cut off
by, say, the spacing between microorganisms), one can
see from the profiles of the plots of Eq. (\ref{eqn: 2dGF}) in Fig. $\ref{fig:
gf2d}$ that the overall response amplitude is again suppressed at strong
advection. Fig.
$\ref{fig: gf2d}$ also illustrates   the process of symmetric  transverse
contraction, in addition to the  asymmetric downstream elongation
and upstream contraction in the longitudinal direction.

\section{Longitudinal striations in the steady state  for
strong advection in $d \ge 2$ }\label{sec:
striation}
Although the linear response theory developed in
Sec. \ref{sec: linear_response} suggests that the transverse localization
length of
an isolated growth hot spot shrinks to zero  as the advection speed increases,
we now
show
that superposition of the responses from uncorrelated growth disorder leads
to correlations  whose transverse correlation length approaches
a finite velocity-independent
value, while the longitudinal correlation length grows linearly with
$\tilde \vel.$ The result is striated population correlations in the steady
state. 
\begin{figure}[htbp]
\begin{minipage}[t]{0.5\linewidth}
    \includegraphics[width=\linewidth]{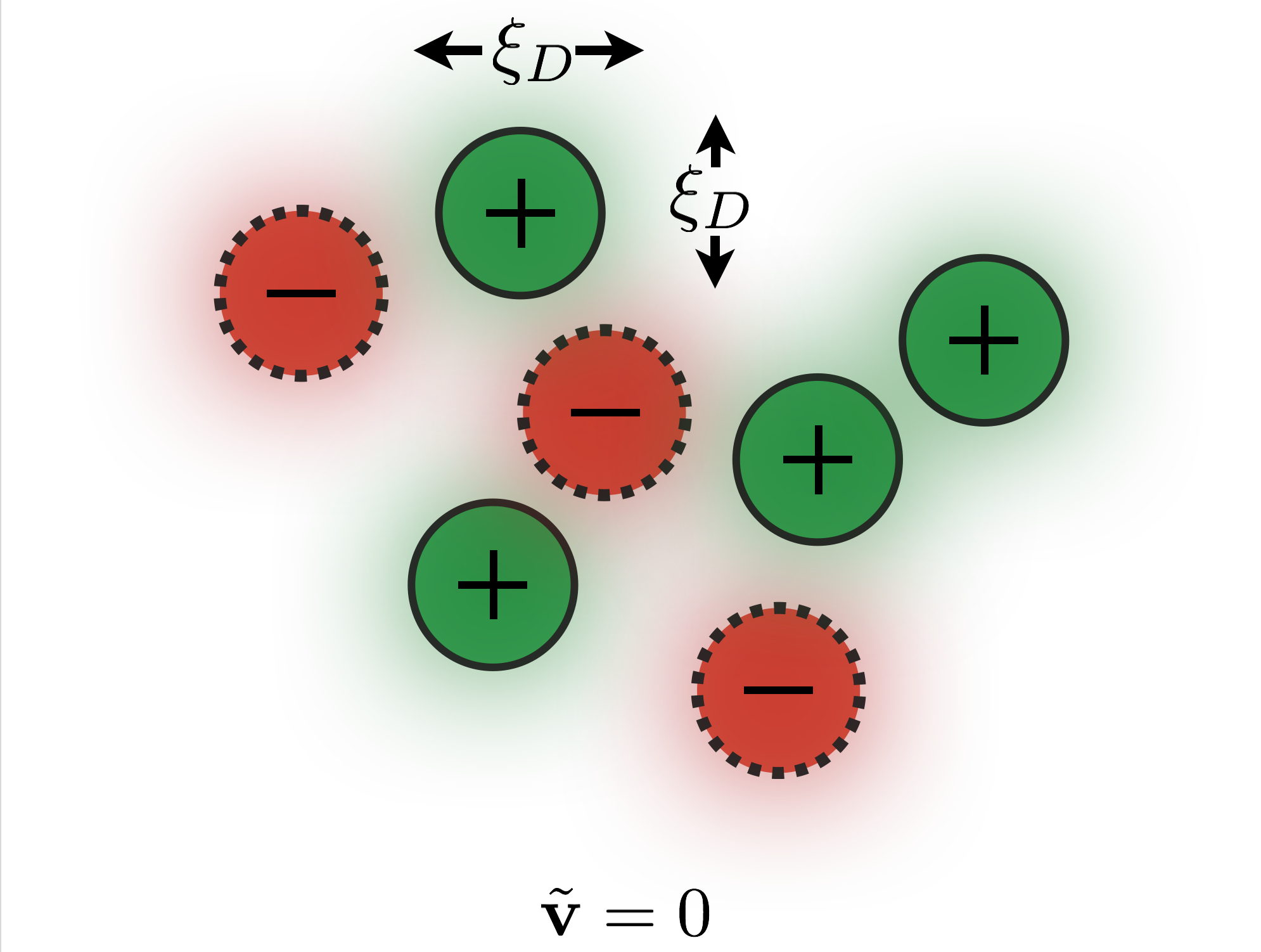}
    \label{f1}
\end{minipage}%
    \hfill%
\begin{minipage}[t]{0.5\linewidth}
    \includegraphics[width=\linewidth]{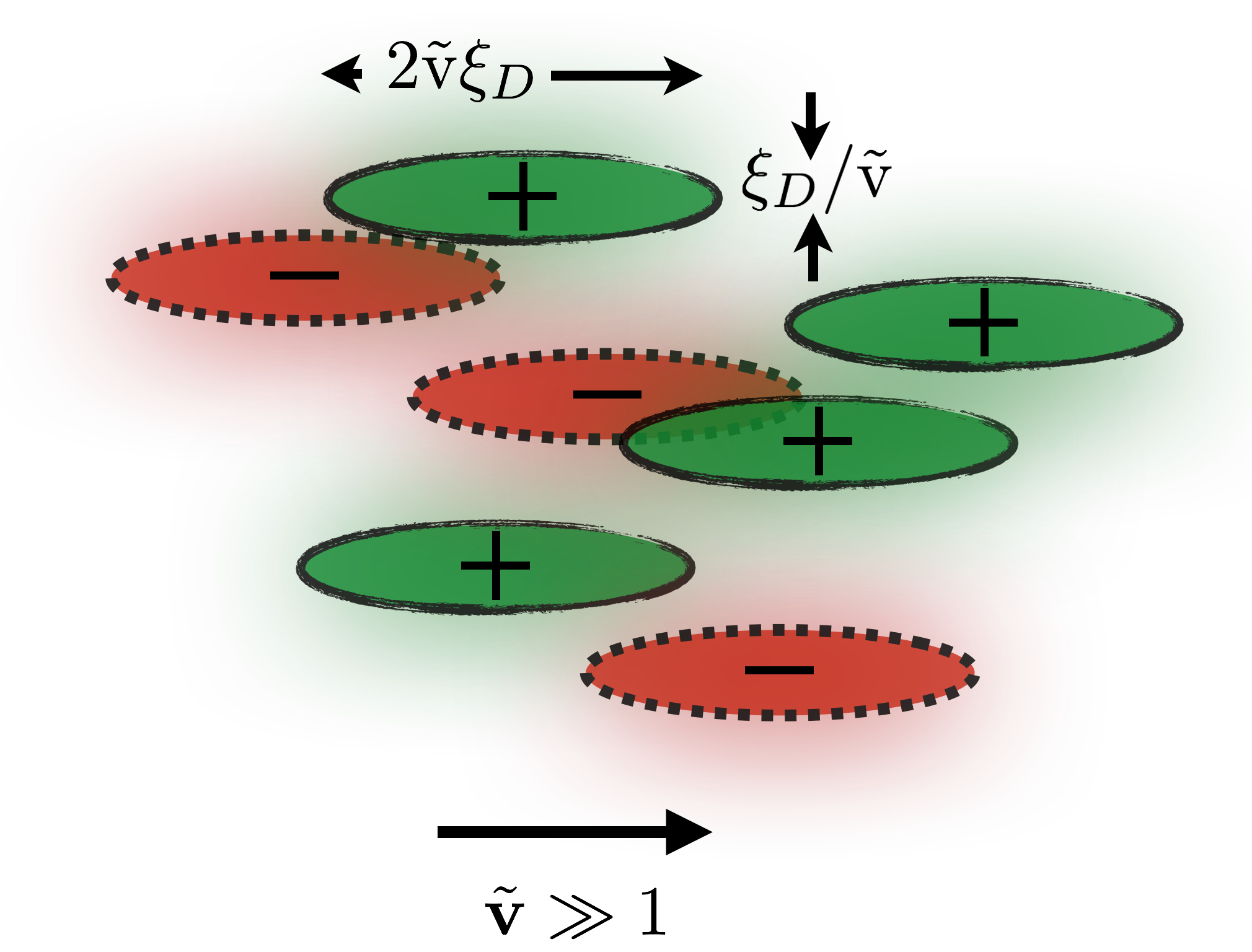}
    \label{f2}
\end{minipage}
\caption{(Color online) How striations driven by flow and spatially quenched
growth disorder
arise in two dimensions. (Left) Without advection, density fluctuations arise
from the  superposition
of the
isotropic responses $G_{\vel=0}(\boldsymbol x-\bx')$ centered  around a source
($U(\boldsymbol x') >0$ ) or  a sink ($U(\boldsymbol
x') <0$ ), depicted in green and red respectively.  The localization length
is given by  $\xi_D
= \sqrt{D/a_0.}$ (Right) For  strong
advection $\tilde \vel \gg 1$, each response  is distorted  asymmetrically:
the downstream  localization length grows  as $\xi_\parallel^+ = 2\tilde \vel
\xi_D,$ and the transverse localization length contracts as $\xi_\perp =
\xi_D/\tilde \vel.$ Although each  response is subjected to transverse contraction
that  shrinks to zero at stronger flow, the random superposition of responses
exhibit
 non-vanishing transverse correlations, as shown in Figs. \ref{fig: isotropy_2d}
and \ref{fig: 2d_structurefactor}. } 
\end{figure}

\subsection{Two-point correlation function and the striation pattern}\label{sec:
two-point corfunc}
To model  rapid, statistically isotropic spatial variations in the growth
rates, we assume each local growth rate $U(\bnd{x})= \delta a(\bnd{x})/a_0$
is drawn from a uniform
box distribution in the interval   $ [-\Delta/a_{0}, \Delta/a_0], $ which
gives the two-point noise correlation function 
\begin{equation}\label{eqn: noise_cor_x}
\langle
U(\bnd{x})U(\bnd{x}')\rangle = \frac{
1}{3}\frac{\Delta^2}{a_0^2}\delta^d(\bnd{x} - \bnd{x}'),
\end{equation}
where $\mean{\cdot}$ denotes ensemble averages over disordered growth rate
realizations.
In the Fourier domain, Eq.(\ref{eqn: noise_cor_x}) gives white noise
with
magnitude\begin{equation}\label{eqn: whitenoise_U}
\langle |U(\bnd{q})|^{2}\rangle= \frac{1}{3}\frac{\Delta^2 }{a_0^2}.
\end{equation}

We expect
statistical translational invariance, $  \langle \phi(\bnd{x})
\phi({\bf 0}) \rangle \equiv  \langle \phi(\bnd{x}+\bnd{x}')
\phi(\bnd{x} ') \rangle,   $ so the two-point correlation function of the
density
fluctuations is given by
\begin{equation}
\langle \phi(\bnd{x})
\phi({\bf 0}) \rangle = \int
\frac{d^d \tilde q}{(2\pi)^{d} }
e^{i\bnd{q} \cdot \bnd{x}}S(\bnd{q}),\label{eqn: 2pts_cfn_nd}
\end{equation}
where the static structure factor in the steady state is related to the Fourier
transformed Green's function $G(\tbq)$,
\begin{align}
S(\bnd{q}) &\equiv \langle |\phi(\bnd{q})|^2
\rangle \nonumber\\
&=\langle |U(\bnd{q})|^2\rangle |G(\bnd{q})|^{2}\nonumber\\
&=\frac{\langle |U(\bnd{q})|^2\rangle}{\left[(\tilde q^2+1)^{2}+4 \tilde\vel^{2}
\tilde q^{2}_\parallel \right ]}. \label{eqn: sfactor_nd}
\end{align}
Upon substituting Eqs. (\ref{eqn: sfactor_nd}) and (\ref{eqn: whitenoise_U})
into
Eq. (\ref{eqn: 2pts_cfn_nd}),  the two-point correlation function of the
density fluctuations  with spatially uncorrelated random growth rates reads

\begin{equation}\label{eqn: two-point-corfunc_real}
\langle \phi(\bnd{x})
\phi({\bf 0}) \rangle =\frac{\Delta^2}{3a_0^2 } \int \frac{d^d
\tilde q}{(2\pi)^d}
\frac{e^{i\bnd{q} \cdot \bnd{x}}}{\left[(\tilde q^2+1)^{2}+4 \tilde\vel^{2}
\tilde q^{2}_\parallel \right ]}.
\end{equation}

We now evaluate Eq. (\ref{eqn: two-point-corfunc_real}) in various dimensions,
with details relegated to  Appendix \ref{sec: detailed_calculation}.
For $d=1,$ Eq. (\ref{eqn: two-point-corfunc_real})
can be evaluated via contour integration, with the result (setting $\txpar
= \tx$) 
\begin{align}\label{eqn: 1D_CorLength}
\langle\phi(\tilde x)\phi(0) \rangle &=\frac{ \Delta^2
}{24a_{0}^2} \frac{1}{\tilde
\vel\sqrt{1+\tilde \vel^2}}\nonumber\\
& \times\left[\frac{1}{g_-(\tilde \vel)} e^{-g_-(\tilde \vel)|\tilde
x|}-\frac{1}{g_{+}(\tilde \vel)}e^{-g_{+}(\tilde
\vel)|\tilde x|} \right],
\end{align}
where the two exponential decays are controlled by $g_{\pm}(\tilde \vel)
\equiv \sqrt{1+\tilde
\vel^2} \pm \tilde \vel.$  Although this result appears singular at $\tilde
\vel=0,$ the limit $\tilde \vel \rightarrow 0$ is in fact well-defined, and
 given
by
\begin{equation}\label{eqn: 1D_CorFunction_v0}
\lim_{\tilde \vel \rightarrow 0 }\mean{\phi(\tilde x) \phi(0)} = \frac{\Delta^2}{12a_0^2}
\left(1+|\tilde x|\right)e^{-|\tilde x|}.
\end{equation}
The exponential localization associated with a single hot spot is broadened
by a factor $(1+|\tilde x|)$;
the  correlation length, however, is the same as the diffusive
localization length $\xi_D$. For strong advection,
$g_-(\tilde
\vel) = 2\tilde \vel+1/2\tilde \vel +O(1/\tilde \vel^3)$
and $g_+(\tilde \vel) = 1/2\tilde\vel +O(1/\tilde\vel^3);$
the first term in the square bracket of Eq. (\ref{eqn: 1D_CorLength})
dominates, resulting in 
\begin{equation}\label{eqn: large_v_1d}
\lim_{\tilde \vel \rightarrow \infty } \langle \phi(\tilde x_\parallel) \phi(0)
\rangle
\sim \frac{e^{-|\tilde
x_\parallel|/2\tilde \vel}}{\tilde \vel}.
\end{equation}

For $d=2,$ the spatial structure of steady state  is embodied in
the longitudinal  correlation function $\langle\phi(\tilde x_\parallel,0)\phi({\bf
0}) \rangle$ and the transverse correlation function 
$\langle \phi(0,\tilde x_\perp)\phi({\bf 0})
\rangle$ that are defined by Eq. (\ref{eqn: two-point-corfunc_real}). In
the absence of advection, the  correlation function is isotropic and is given
by
\begin{equation}
\lim_{\tilde \vel \rightarrow 0}\mean{\phi(\tilde \bx)\phi(\boldsymbol{0})}
\sim  |\tilde \bx|K_1(|\tilde \bx|),
\end{equation}
where $K_{1}(x)$ is the modified Bessel function of a second kind that
decays exponentially at large distance $x $  as    
$K_{1}(x)=\sqrt{\frac{\pi}{2x}}\exp(-x)\left[ 1 + 3/8x
+\mathcal{O}(x^{-2})\right]$ for $x \gg 1$ \cite{Abramowitz:1964aa}.  The
correlation length is thus given by diffusive correlation
length $\xi_D  $ in this limit.  In the strong advection limit,  however,
correlations  become  highly anisotropic:  
\begin{align} 
\lim_{\tilde \vel \rightarrow \infty }\mean{\phi(\tilde x_\parallel,0)\phi({\bf
0})} &\sim \left(\frac{|\tilde x_\parallel|}{2\tilde
\vel}\right)^{1/4}K_{-1/4}\left(\frac{|\tilde x_\parallel|}{2\tilde \vel}\right),\label{eqn: longitudinal_2d_corr}\\
\lim_{\tilde \vel \rightarrow \infty }\mean{\phi(0, \tilde x_\perp)\phi({\bf 0})}  &\sim \frac{e^{-|\tilde x_\perp|}}{ \tilde \vel}, \label{eqn: transverse_2d_corr}
\end{align}
with the longitudinal and transverse correlation lengths that behave as follows,
\begin{align}
 \lim_{\tilde \vel \rightarrow \infty }\xi_\parallel &=\xi_\vel = \frac{\vel
}{a_0},\\
\lim_{\tilde \vel \rightarrow \infty }\xi_\perp &= \xi_D= \sqrt{\frac{D}{a_0}}.\label{eqn:
transverse_corlen_vinfinite}
\end{align} 
 Note the  limiting transverse correlation length is given by a velocity-independent,
diffusive
localization length. However, the  ratio of transverse to longitudinal
correlation lengths becomes highly anisotropic: 
\begin{equation}\label{eqn: ratio_corrlength}
\lim_{\tilde \vel \rightarrow \infty } \left(\frac{\xi_\parallel }{\xi_\perp}\right)=\frac{\vel}{\sqrt{Da_{0}}}=2\tilde
\vel.
\end{equation}
These anisotropic  patterns might arise in natural marine microbial
populations, their Fisher wave spreading velocity is much smaller than oceanic flow
speed; for a motile bacteria, a typical diffusion constant
$D$ is of order
$10^{-5}$ cm$^2$ s$^{-1}$ and the typical doubling time $a_0$ is of order
$10^{-3}$ s$^{-1}$ which gives $\vel_F = 2\sqrt{D a_0} $ of order $1$
$\mu$m s$^{-1} $ \cite{Murray:2002aa}, which is  indeed small compared to  a typical
oceanic current; see Ref. \cite{Neufeld:2009aa} and references therein.

Note  that uncorrelated disorder (viewed as a superposition
of responses from growth hot spots) leads to a non-zero
transverse correlation length, as shown  in Eq. (\ref{eqn:
transverse_corlen_vinfinite}) for two dimensions,  in contrast to the result
for a single point-like growth hot-spot of Sec. \ref{sec: GF}.
 In fact, a finite transverse correlation arises for all $d\ge
2.$ To see this, consider the transverse correlation functions: 
\begin{align}
\langle\phi(0,\tbxper)\phi(\boldsymbol 0) \rangle &\sim \int  \frac{d^{d-1}\tqper}{(2\pi)^{d-1}}
e^{i \tbqper \cdot \tbxper}\int \frac{d\tqpar}{2\pi} \frac{1}{\left[(\tq^2
+1)^{2}+4
\tilde\vel^{2}
\tqpar^{2} \right]}\nonumber\\
&= \pi\int
\frac{d^{d-1}\tqper}{(2\pi)^{d}} \frac{e^{i \tbqper \cdot \tbxper}}{(\tqper^2+1)(\tqper^2+1+\tv^2)^{1/2}}.
\end{align}  
Upon taking the limit  $\tv \gg 1,$ it follows that\begin{align}
\lim_{\tv \rightarrow \infty}\langle\phi(0,\tbxper)\phi(\boldsymbol 0) \rangle
&\sim \frac{\pi}{\tv}\int
\frac{d^{d-1}\tqper}{(2\pi)^{d}} \frac{e^{i \tbqper \cdot \tbxper}}{(\tqper^2+1)},\label{eqn:
}\\
&= \frac{e^{-|\txper|}}{4\tv},\label{eqn: transv_corlen}
\end{align}
which show that the transverse correlation length still obeys Eq.
(\ref{eqn: transverse_corlen_vinfinite})
for $d\ge 2$. Thus, the limiting
transverse correlations in higher dimensions
remains \textit{velocity-independent} and is characterized by the diffusive
localization length $\xi_D$, while the amplitude of fluctuations decays at
increasing advection as $\lim_{\tv \rightarrow \infty}\langle\phi^{2}(\boldsymbol
0) \rangle \sim 1/\tilde \vel. $

\subsection{Comparisons to numerical simulations}\label{sec: numerical_sim}
We now check theoretical predictions in Sec. \ref{sec: two-point corfunc} by
comparing the theoretical structure factor $S_{\mathrm{th}}(\bnd{q})$ given
by Eq. (\ref{eqn: sfactor_nd}) to the  structure factor $S_\mathrm{num}(
\bnd{q}) $  calculated from the numerical solution of Eq. (\ref{eqn: SteadyState}).
To obtain the numerical steady state satisfying Eq. (\ref{eqn: SteadyState}),
we inoculate the simulation domain with a uniform concentration equal to the mean carrying
capacity  $\bar K, $ and numerically  evolve the  time-dependent advective FKPP Eq. (\ref{eqn:
Gen_FKPP}) until the steady state  is reached using
the  Lattice Boltzmann method, a robust lattice discretization scheme
for advective and diffusive transports \cite{Cali:1992aa,Succi:2015aa} that can incorporate
reactive agents \cite{Succi:2001aa}. This method  achieves
high numerical accuracy for reaction-diffusion-advection problems, even
in the strong advection limit \cite{Succi:2014aa}.
 Random growth rate   $U(\bnd{x}_i)=
\delta a(\bnd{x}_i)/a_0$ is introduced on each lattice site $\bnd{x}_i,$
and is independently
drawn from a uniform
box distribution in the interval   $ [-\Delta/a_0, \Delta/a_0]. $ The
9-speed 2-dimensional
(D2Q9) lattice is adopted to evolve the density  field according to   Eq.
(\ref{eqn:
Gen_FKPP}) on a two-dimensional square lattice with $N_{x} \times N_y$ sites
and a periodic boundary condition  implemented by  adding the buffer  sites
at the boundaries \cite{Succi:2001aa}. In this scheme, time evolution on
a one-dimensional lattice with a periodic boundary condition also follows
immediately, provided $N_{x} $ is set to 1. Once
the numerical steady state $c_{\mathrm{num}}^*(\bx)$ is reached, one can
calculate
the squared modulus of the discrete fourier transform of the steady state
density fluctuations,  denoted by $|\phi_{\mathrm{num}}(\tbq)|^2$. After
taking ensemble averages
over the random growth rates, this results in the numerical structure factor
 $S_\mathrm{num}(
\bnd{q}). $  This numerical structure factor is the discrete  counterpart
of the theoretical structure factor   $S_{\mathrm{th}}(\bnd{q})$
predicted by  
 Eq. (\ref{eqn: sfactor_nd}). 

\begin{figure}[!h]     
\includegraphics[width =0.7\linewidth]{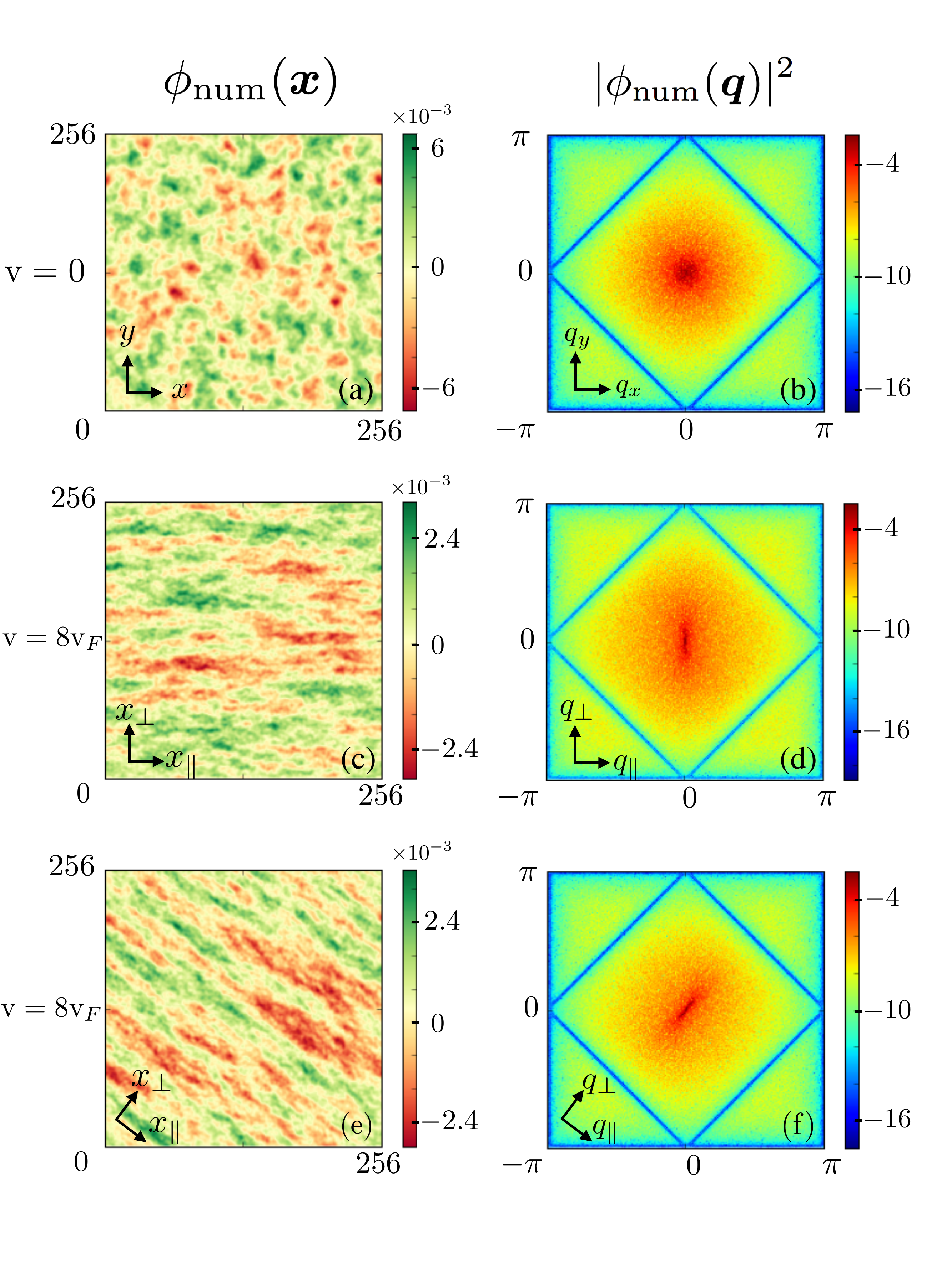}
\caption{
(Color online) Steady state density fluctuations $\phi_{\mathrm{num}}(\bx)
\equiv [c_{\mathrm{num}}^*(\bx) - \bar K]/\bar K$ from Lattice Boltzmann
simulations with $256^2$ sites (left) and  the squared
modulus of the discrete fourier transform of the steady state
density fluctuations  $|\phi_{\mathrm{num}}(\bq)|^2$ (right) for the same
disorder realization but different flow velocities: $\vel =  0$
in (a) and (b), $ \vel = 8 \vel_F$ along
the $x$-axis in (c) and (d), and   $\vel = 8\vel_F$ inclined at a  $37^{\mathrm{o}}$ angle relative to the periodic boundary conditions
in  (e) and (f). The coordinate $\bx =(x,y)$
and the wavevector $\bq = (q_{x},q_{y})$ are displayed in the original
(dimensional)
scale with the lattice unit  $l_0$  set to 1. The color codes for the structure
factors ((b), (d), and (f))  are displayed in base-10 logarithmic scale.
The parameters
in this figure
are $D =5\times10^{-4}, a_0
= 10^{-5}, \Delta = 0.1a_0,$ and  $b = 10^{-9},$ which give
$\xi_D = \sqrt{D/a_0}\approx7,$  $\vel_F = 2\sqrt{D a_0} \approx
1.4\times 10^{-4}$ , and $\bar K=a_0/b = 10^{4}$.
 Long wavelength modes in the absence
of advection, in (b), appear statistically isotropic as expected. The blue diamonds at large $q$ in (b), (d), and (f) are an artifact of the underlying lattice.}   
\label{fig: isotropy_2d}
\end{figure}
 
\begin{figure}[!ht]
\includegraphics[width =0.7\linewidth]{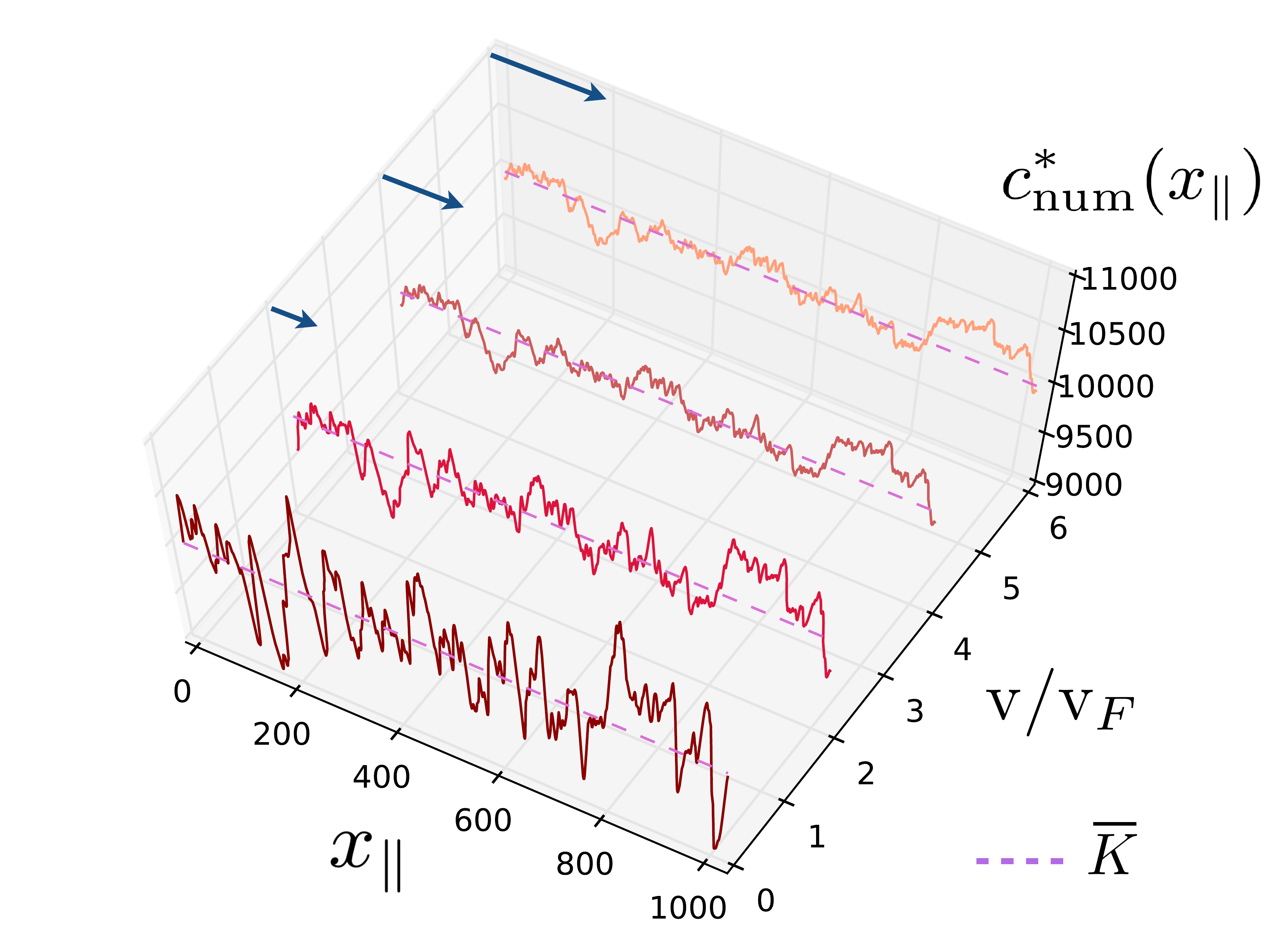}\caption{
Typical steady-state densities $c_{\mathrm{num}}^*(x_\parallel)$ in \textit{one
dimension} due to growth
disorder and advection from the lattice Boltzmann simulation using a $D2Q9$ scheme
of size $1\times 1024$ sites.   The parameters are $D =5\times10^{-4}, a_0
= 1\times10^{-5}, \Delta = 0.4 a_0,$ and  $b = 10^{-9},$ resulting in
$\xi_D = \sqrt{D/a_0}\approx7.07,$ $\vel_F = 2\sqrt{D a_0} \approx
1.4\times 10^{-4}$ , and $\bar K=a_0/b =10^{4}$. The dashed  lines denote the local carrying capacity $\bar K$. As advection becomes stronger,   density fluctuations are   suppressed and (longitudinal) correlation lengths stretch out. The details of correlations in one dimensional systems are elucidated in the structure factors of Fig.\ref{fig: 1dsfactor}.}
\label{fig: 1D_xi_pm}         
\end{figure}

Figures \ref{fig: 1D_xi_pm} and \ref{fig: 2dsteadystates} report numerical
steady states obtained from the Lattice Boltzmann simulations with random
growth rates in one and two dimensions. These simulations show  the decrease
in the amplitude of the density fluctuations as advection becomes stronger,
confirming the assumption of the linear response theory in Sec. \ref{sec:
linear_response}. In addition, unlike  standard  finite-difference schemes
in which advection can lead to spurious alignment of population structure
along  lattice directions  \cite{Dahmen:1999aa}, the D2Q9 discretization, which ensures
the fourth order isotropy of lattice tensors,  avoids
the artifact of preferred lattice orientations. The left column of Fig. \ref{fig:
isotropy_2d} illustrates the steady state density fluctuations for the same
realization of the random growth rate for different advection velocities:
 $\boldsymbol \vel = \boldsymbol 0$ in Fig. 5(a), $ \vel = 8 \vel_F$ along
the $x$-axis in Fig. \ref{fig: isotropy_2d}(c), and   $\vel = 8 \vel_F$ inclined at $37^{\mathrm{o}}$
relative to the directions associated with the periodic boundary conditions in  Fig. \ref{fig: isotropy_2d}(d). The steady state populations in Figs. \ref{fig: isotropy_2d}(c) and (e) streak out along the advection
direction, despite the bias imposed by an underlying  square lattice. The elongation of steady
state populations along the advection direction  also appears in the Fourier-transformed
$|\phi_{\mathrm{num}}(\bq)|^2$  shown on the right column of Fig. \ref{fig:
isotropy_2d}. There, the short wavelength modes in the longitudinal direction
are noticeably suppressed, while the modes in the transverse direction are
only mildly modified, leading to  elliptical contours  at small $q$'s with
the minor and major axis aligned along the  longitudinal and transverse direction,
respectively. Notice the plots of   $|\phi_{\mathrm{num}}(\bq)|^2$ are approximately
zero  whenever  $\bq \cdot \boldsymbol{r}_{\mathrm{lat}} = \pm\pi/l_{0} =
\pm \pi,$ where $ \boldsymbol{r}_{\mathrm{lat}}$ belongs to either the nearest
neighbor   basis  $ \{ \boldsymbol e_x,  \boldsymbol e_y\}$  or the next nearest neighbor
basis $\{\boldsymbol e_x + \boldsymbol e_y, \boldsymbol e_x - \boldsymbol e_y\}$  of the square lattice,
corresponding to the 
blue squares or the  blue diamonds, respectively.  These
short-wavelength anisotropic lattice artifacts  arise from the    D2Q9 scheme
that allows both  nearest and  next nearest neighbor hopping on the square
lattice. However, the long-wavelength modes ($q$ near the origin), which
characterize the macroscopic    striations, are further away from the  blue diamond
and are 
orders of magnitude larger than zone-boundary modes  with wavenumber $q \gtrsim \pi/\sqrt{2}l_0= \pi/\sqrt{2}$.
 As discussed above, the long-wavelength physics is insensitive to these lattice artifacts, as can be confirmed by rotating Fig. \ref{fig:
isotropy_2d}(f) counterclockwise by an angle $37^{\mathrm{o}}$ and comparing
to Fig. \ref{fig:
isotropy_2d}(d). We thus expect that the long-range
striation patterns are well-described  by the behavior
of $S_\mathrm{num}(
\bnd{q})$  near the origin, as determined by the Lattice Boltzmann simulations, averaged
over ensembles of growth rates.

We now compare the long-wavelength modes of $S_\mathrm{num}(
\bnd{q})$  to  $S_{\mathrm{th}}(\bnd{q})$
predicted by 
 Eq. (\ref{eqn: sfactor_nd}). Although theoretical results of Sec. \ref{sec:
two-point corfunc} assume
spatially uncorrelated growth disorder whereas the Lattice Boltzmann simulation
has
an intrinsic short-range disorder correlation on the order of the lattice
size $\l_0\equiv 1$, we expect that the  long-distance statistics are insensitive to such microscopic details, provided
the correlation length of the emerging pattern is larger than   the
lattice size.  Because advection stretches out correlations in the longitudinal direction, the approximation to a continuum model should become even more accurate at stronger advection. In fact, as shown
in the comparisons between 
$S_\mathrm{num}(\bnd{q})$  to  $S_{\mathrm{th}}(\bnd{q})$ in one and two
dimensions in Figs  $\ref{fig:
1dsfactor}$ and \ref{fig: 2d_structurefactor}   respectively,
excellent agreement  is obtained at long wavelengths  even when  $\xi_D$
 is of order 2$l_0 = 2$, which is only twice the  correlation length of the
simulated noise.      The agreement is even better for
stronger advection.  Thus, the uncorrelated noise predictions
of Sec. \ref{sec:
two-point corfunc} are able to capture the long-range statistics of the striation
pattern, despite the presence of inherent lattice-scale  correlations of the
simulated noise.

\begin{figure}[!h]     
\includegraphics[width =0.7\linewidth]{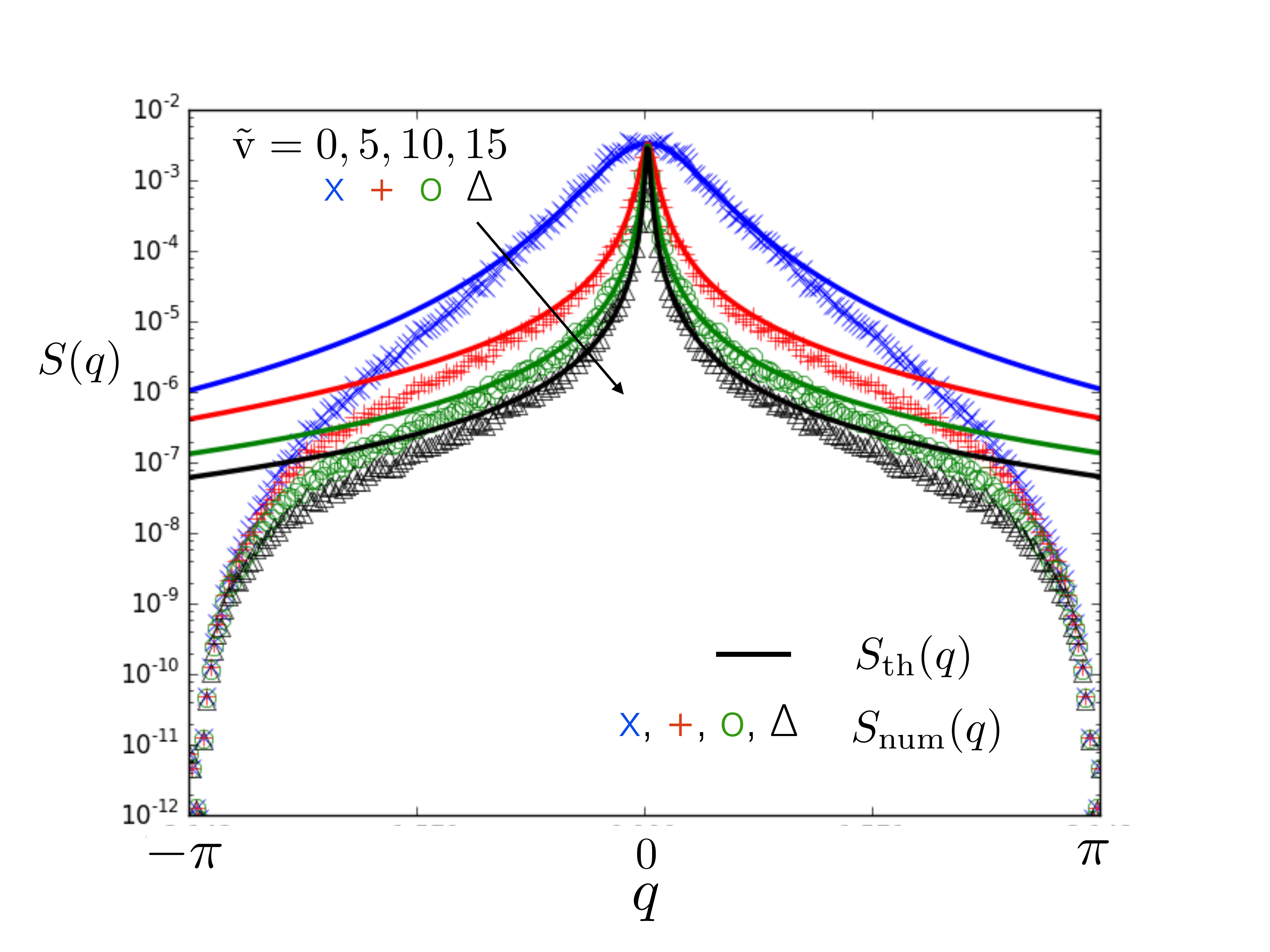}\caption{
(Color online) Comparisons between the predicted \textit{one-dimensional}
structure
factor  $S_{\mathrm{th}}(q)$ (solid lines),  and  numerical structure factor
$S_{\mathrm{num}}(q)$  (symbols) from the lattice Boltzmann simulation with
$1\times256$  sites. The wavenumber $q$ is displayed in the original (dimensional)
scale with the lattice unit  $l_0$  set to 1. Simulation results are obtained
from averaging the structure
factors of density fluctuations over 200 growth rate realizations. The predicted
long-wavelength modes of  $S_{\mathrm{th}}(q)$ are in excellent agreement
with
those of the simulated 
$S_{\mathrm{num}}(q)$. Although the simulated structure factors are affected
by inherent lattice-size disorder correlations near $q \sim \pi/l_0 = \pi$,
 the $q \rightarrow 0$ modes are much larger  (note the logarithmic scale
on the $S(q)$ axis). At strong advection when $ \tilde \vel \gg 1$, long-wavelength
modes are even more pronounced, and  $S_{\mathrm{th}}(q) $ is even closer to $S_{\mathrm{num}}(q).$ The parameters in this figure are $D =5\times10^{-4},
a_0
= 9\times10^{-5}, \Delta = 0.1 a_0,$ and  $b = 10^{-9},$ giving 
$\xi_D = \sqrt{D/a_0}\approx2.36,$ $\vel_F = 2\sqrt{D a_0} \approx
4.2\times 10^{-4}$ , and $\bar K=a_0/b =9 \times 10^{4}$.
}
\label{fig: 1dsfactor}
\end{figure}

\begin{figure*}[ht!]
\includegraphics[width = \linewidth]{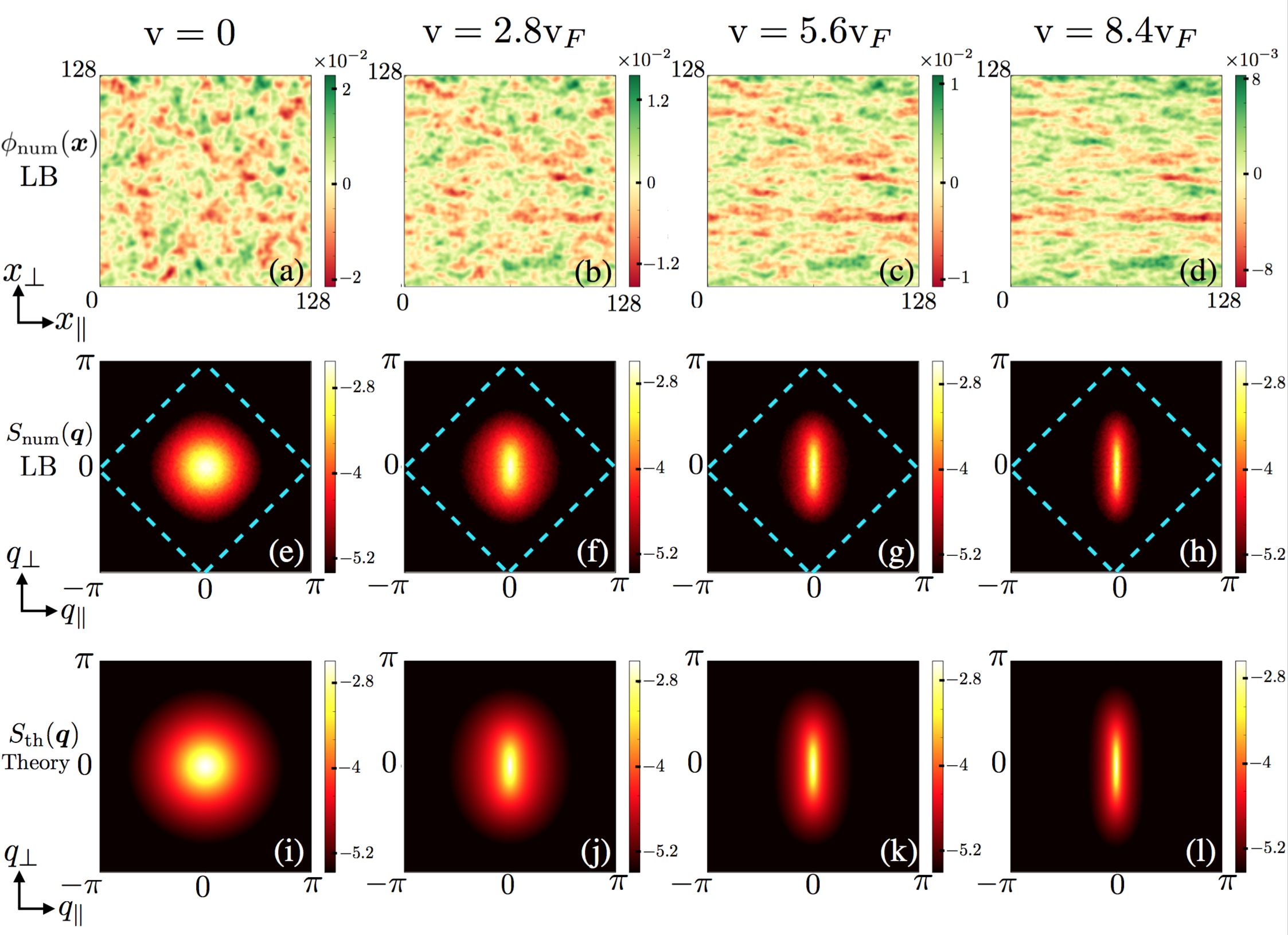}\caption{
(Color online) Comparisons between the predicted \textit{two-dimensional}
structure
factor $S_{\mathrm{th}}(q_\parallel,q_\perp)$ in the third row ((i)-(l))
and the simulated structure factor
$S_{\mathrm{num}}(q_\parallel,q_\perp)$  in the second row ((e)-(h)) from
the Lattice Boltzmann (LB) simulations with
$128^2$  sites. Typical  simulated steady-state density fluctuations $\phi_{\mathrm{num}}(x_\parallel,x_\perp)$
 from the same disorder realization are shown in the first row ((a)-(e))
. The coordinate $\bx =(x_\parallel,x_\perp)$
wavevector $\bq = (q_\parallel,q_\perp)$ are displayed in the original
(dimensional)
scale with the lattice unit  $l_0$  set to 1. Simulation results are obtained
from averaging the structure
factors of density fluctuations over 200 disorder realizations.  Color codes
for the structure factors ((e)-(l))  are displayed in base-10 logarithmic
scale. To highlight the agreements
between the predicted long wavelength modes and the simulated ones, the
values  below $10^{-3}$-times  the maximum are all represented in black; the
predicted
long-wavelength modes of  $S_{\mathrm{th}}(q)$ are in excellent agreement
with
those of the simulated 
$S_{\mathrm{num}}(q)$. The next-nearest neighbor effect of the D2Q9 scheme
represented
by the dashed blue diamonds in $S_{\mathrm{num}}(q_\parallel,q_\perp)$ is
similar
to that displayed in  Fig. \ref{fig: isotropy_2d}, and
is many order of magnitude smaller than the long-wavelength modes. At strong
advection (right columns), the longitudinal long-wavelength
modes are even more pronounced while the transverse long-wavelength modes
are not significantly affected; this longitudinal compression in the structure
factors indicates\ the formation of striated population structure exemplified
in (c) and (d). The parameters in this figure are
$D
=5\times10^{-4}, a_0
= 9\times10^{-5}, \Delta = 0.1a_0,$ and  $b = 10^{-9},$ which give
$\xi_D = \sqrt{D/a_0}\approx2.36, $  $\vel_F = 2\sqrt{D a_0} \approx
4.2\times 10^{-4}$, and $\bar K=a_0/b =9 \times 10^{4}$.
} 
\label{fig: 2d_structurefactor} 
\end{figure*}

\section{Conclusions and outlook}\label{sec: conclusion}

 Previous work has demonstrated the important role of spatially quenched disorder
in the local growth rates on the long-time population
structure 
that disperses through unidirectional advection and diffusion in  hostile
growth environments
 \cite{Dahmen:2000aa, Lin:2004aa, Neicu:2000aa, Nelson:1998aa}. Here, we
study the role of growth disorder on the long-time population structure of
populations in \textit{fertile} random growth environments in which the local
carrying
capacity is strictly positive with weak disorder in the growth rates. To
leading order in the strength of weak growth disorder, fluctuations  from
the homogeneous carrying capacity of  steady-state
populations can be perturbatively treated as a  response to growth disorder.

Within the linear response theory, growth disorder and strong fluid advection
leads to the emergence of striated steady-state population structure in two
or higher dimensions with a periodic boundary condition. For large advection speeds $ \vel \gg \vel_F = 2\sqrt{D a_0}$, the longitudinal correlation
length elongates as  $\vel/a_0$ and the transverse correlation length
approaches the finite velocity-independent value given by the diffusive
localization length 
$ \xi_D = \sqrt{D/a_0}.$ 
 In contrast to steady-state populations with  
a point-like growth hot spot, whose transverse localization length contracts
to zero as the advection speed increases, spatially quenched random growth
disorder  impedes the disappearance of transverse correlations.
In our case, transverse correlations    exhibit anomalous transverse diffusive spreading
behavior of the early-time growth dynamics at strong advection as studied
in Ref. \cite{Nelson:1998aa}. 
The Lattice Boltzmann simulation of the  generalized FKPP equation (\ref{eqn:
Gen_FKPP})  confirms the emergence of striated population structure  at strong
advection, and verifies the predicted long-distance statistics from the linear
response theory in one and two dimensions. Similar striated population structures might arise in  natural marine microbial
populations in spatially disordered growth environments
and transported along   closed recirculating flows,  such as oceanic flows
characterized by coherent Lagrangian vortices discussed in Refs. \cite{Haller:2013aa,Haller:2015aa}.

In two-dimensional open flows, where populations do not actually recirculate, we  numerically observe (not
shown in this paper) the pinning-depinning phenomena
of the population frontier, the interface between occupied and unoccupied
regions,  analogous to those observed in autocatalytic chemical reaction
front propagating in a porous media with a background fluid flow \cite{Atis:2013aa,Saha:2013aa,Gueudre:2014aa}.
 In particular, the population frontier is pinned 
by  growth rate disorder despite the presence of  advection up to the critical
advection speed, above which  populations are completely washed away from
the domain of interest. It would be interesting to explore, using the Lattice Boltzmann method,  whether this
pinning-depinning
transition is a dynamic critical phenomena and, if so, whether
it belongs to the  universality class studied in Refs. \cite{Atis:2013aa,Saha:2013aa,Gueudre:2014aa}.

\begin{acknowledgments}
We thank S\'everine Atis for  useful discussions on pinning-depinning phenomena.
This work was supported
in part by the National Science Foundation (NSF) through Grant No. DMR-1306367 and by the Harvard
Materials Research Science and Engineering Laboratory, through Grant No. DMR-1420570.
Portions of this research were conducted during a stay at the Center for
Models of Life at the Niels Bohr Institute, the University of Copenhagen.
Computations were performed on the Odyssey cluster supported by the FAS Division
of Science Research Computing Group at Harvard University. S.S. was supported
by the Institute for Applied Computational Science
at John A. Paulson School of Engineering and Applied Sciences at
Harvard University, and by the 
Integrated Mesoscale Architectures for Sustainable Catalysis 389 (IMASC)
Energy Frontier Research Center (EFRC) of the Department of Energy, Basic
Energy Sciences, Award number 390 DE-SC0012573.  

\end{acknowledgments}

\appendix
\section{Green's Functions, correlation functions, and
structure factors}\label{sec: detailed_calculation}
 This appendix contains calculation details of analytical results obtained
in this paper.
First, we solve  Eq. (\ref{eqn: GF_EqnFor_non-d})
by
a direct Fourier transformation, as a check on
Eq. (\ref{eqn: GF_gauged_relation}). Upon setting $(\Delta /a_0)$ to
unity for convenience, the Green's function of Eq. (\ref{eqn: GF_EqnFor_non-d})
 in Fourier-space reads  \begin{equation}\label{eqn: fourier_mode}
G(\bnd{q})=\frac{1}{\tilde q^2-2i\tilde\vel \tilde q_\parallel +1}.
\end{equation} 
Inverse fourier transformation gives the real-space Green's function in $d$-dimensions: 
\begin{equation}\label{eqn: IFT}
G(\tbx) = \int \frac{d^d\tilde q}{(2\pi)^d}
\frac{e^{i\bnd{q} \cdot \bnd{x}}}{\tilde q^2-2i\tilde \vel \tqpar + 1  }.
\end{equation}
To evaluate (\ref{eqn: IFT}), we use a  trick, often employed to calculate correlation functions in field theory 
\cite{Zeidler:2008aa}, to convert
the denominator into an exponential integral: \begin{equation}
\frac{1}{X^n}=\int_0^\infty ds \ \frac{s^{n-1} e^{-sX}}{(n-1)!}.
\end{equation}
With $n=1$, (\ref{eqn: IFT}) becomes  
\begin{align}
G(\tbx) &= \int_0^\infty ds \int \frac{d^d\tq}{(2\pi)^d}
e^{i\tbq\cdot \tbx} e^{-s(\tq^2-2i\tv\tqpar +1)},\nonumber\\
&=  \int_0^\infty ds \ e^{-s
 } \underbrace{\left(\int \frac{d^{d-1}\tqper}{(2\pi)^{d-1}}e^{i\tbqper \cdot
\tbxper-s\tqper^2}\right)}_{\frac{1}{(2\pi)^{d-1}}\sqrt{\frac{\pi}{s}}^{d-1}\exp({-\txper^2/4s})}\nonumber\\
&\hspace{2cm}\times\underbrace{\left( \int \frac{d\tqpar}{(2\pi)}e^{i\tqpar
(\txpar
-2s\tv)-s\tqpar^2} \right)}_{\frac{1}{(2\pi)}\sqrt{\frac{\pi}{s}}\exp[{-(\txpar-
2s\tv)^2/4s}]},\nonumber\\
&=      \frac{\exp({\tv \txpar) }}{(2\pi)^{d/2}}
\int_0^\infty
ds \ s^{-d/2}e^{-s\left(1 +\tv^2\right)-\tx^2/4s}. \label{eqn:
gf_calculation}
\end{align} 
Upon recalling the integral representation  of the modified Bessel
function of a second kind \cite{Olver:2010aa}
\begin{equation}
K_\nu(z) = \frac{1}{2}\left(\frac{1}{2}z\right)^\nu\int_0^\infty dt  \ t^{-\nu
-1}e^{-t-z^2/4t},
\end{equation}
and using the symmetry   $K_{-\nu}(z) = K_{\nu}(z),$ 
(\ref{eqn: gf_calculation}) becomes
\begin{equation}
G(\tbx)=\frac{\exp({{\tv \txpar }) }}{(\pi)^{d/2}}\left[\left(\frac{|\tbx|}{\sqrt{1+\tilde
\vel^2}}\right)^{1-d/2}
K_{1-d/2}\left(\sqrt{1+\tilde \vel^2}|\bnd{x}|\right)\right],\label{eqn:
GF_nd}
\end{equation}
which is identical to Eq. (\ref{eqn: GF_gauged_relation}) with the gauge
transformed Green's function $G_{\tv}(\tbx)$ of Eq. (\ref{eqn: GF_dimensionless_gauged})
 by the square bracket (after restoring the factor $\Delta
/a_0$).  

We now outline the details of calculations to achieve analytical results
in Sec. \ref{sec: two-point corfunc} that follow from the general prescription
of the two-point correlation function of Eq. (\ref{eqn: two-point-corfunc_real}),
where again we set $\Delta/a_0 = 1$ for short. For $d=1$, the correlation
function reads
\begin{equation}\label{eqn: contour}
\langle\phi(\tilde x)\phi(0) \rangle = \frac{1}{6 \pi}\int_{-\infty}^{\infty}
d\tilde q \ e^{i\tilde q \tilde x}
\displaystyle{\prod_{k=1}^4}\frac{1}{ (\tq - \tq_{k})},
\end{equation} 
where the four poles, that arise from the factorization of 
$\left[(\tilde q^2+1)^{2}+4 \tilde\vel^{2}
\tilde q^{2} \right],$ are located at 
\begin{align*}
\tilde q_1 &= -\tq_2 =i(\sqrt{1+\tilde\vel^2} +\tilde\vel ),\\
\tilde q_3 &= - \tq_4 = i(\sqrt{1+\tilde\vel^2} -\tilde\vel).\\
\end{align*} 
Closing the contour in the upper (lower) half plane for $\tilde x > 0 $ $(\tilde
x < 0),$ then using the residue theorem yields the exact result of the one-dimensional
correlation function given by Eq. (\ref{eqn: 1D_CorLength})

For $d=2,$ we first calculate the transverse two-point correlation function:
\begin{align}
\langle\phi(0,\txper)\phi(\boldsymbol 0) \rangle &\sim \int  \frac{d\tqper}{2\pi}
e^{i \tqper \txper}\int \frac{d\tqpar}{2\pi} \frac{1}{\left[(\tq^2 +1)^{2}+4
\tilde\vel^{2}
\tqpar^{2} \right]}\nonumber\\
&= \frac{1}{4\pi}\int
d\tqper \frac{e^{i \tqper \txper}}{(\tqper^2+1)(\tqper^2+1+\tv^2)^{1/2}}.
\end{align}
In the limit $\tv \rightarrow \infty,$ we obtain 
\begin{align}
\lim_{\tv \rightarrow \infty}\langle\phi(0,\txper)\phi(\boldsymbol 0) \rangle
&\sim \frac{1}{4\pi\tv}\int
d\tqper \frac{e^{i \tqper \txper}}{(\tqper^2+1)},\label{eqn: }\\
&= \frac{e^{-|\txper|}}{4\tv}.
\end{align}

Now consider the longitudinal two-point correlation function.
\begin{align}
\langle\phi(\txpar,0)\phi(\boldsymbol 0) \rangle &\sim \int  \frac{d\tqpar}{2\pi}
e^{i \tqpar \txpar}\int \frac{d\tqper}{2\pi} \frac{1}{\left[(\tq^2 +1)^{2}+4
\tilde\vel^{2}
\tqpar^{2} \right]},\nonumber\\
&=  \int  \frac{d\tqpar}{2\pi}
e^{i \tqpar \txpar}\left\{\frac{1}{8i\tv|\tqpar|}\left[\frac{1}{\sqrt{f_+(\tqpar)}
} - \frac{1}{\sqrt{f_-(\tqpar)}
}\right]\right\}, \label{eqn: appdx_transverse_cal} \\
\end{align}
where the term  in the curly bracket is the result of contour integration
with respect to $\tqper$with the contour in the upper (lower) half plane
for $\tqpar >0$ ($\tqpar <0$), and  $f_\pm(\tqpar)\equiv (\tqpar^2 +1)\pm
2i\tv\tqpar. $ Because the integrand is real, the resulting integral must
be real although it manifestly contains an imaginary part. To rewrite the
squared bracket explicitly as a real-function, consider the following change
of variable:  $re^{i\theta_\pm} \equiv f_\pm(\tqpar).$ Then, $r$ and $\theta_\pm$
are given by $r =f_+(\tqpar)f_-(\tqpar)=(\tqpar^2 +1)^{2}+
4\tv^{2}\tqpar^{2},$ and $\theta_+ = - \theta_-$ with $\tan(\theta_+)=2\tv\tqpar/(\tqpar^2+1).
$
After some algebra and trigonometric identities, one finds the term in the
curly bracket of (\ref{eqn: appdx_transverse_cal}) becomes
\begin{align}
\left[\frac{1}{4i r|\sin(\theta_+)|}\frac{1}{\sqrt{r}}\left(e^{-i\theta_+/2}
-e^{-i\theta_-/2}\right) \right] &= \frac{1}{2\sqrt{2}}\frac{r^{-3/4}}{\sqrt{1+\cos(\theta_+)}}
\end{align}
Substituting the expression into (\ref{eqn: appdx_transverse_cal}), we obtain
\begin{align}
&\langle\phi(\txpar,0)\phi(\boldsymbol 0) \rangle =  \int  \frac{d\tqpar}{4\sqrt{2}\pi}
e^{i \tqpar \txpar}\\ 
&\times \frac{1}
{[(\tqpar^2 +1)^{2}+
4\tv^{2}\tqpar^{2}]^\frac{3}{4}} \frac{1}{\left\{ 1+ (1+\tqpar^2)/[(\tqpar^2
+1)^{2}+
4\tv^{2}\tqpar^{2}]^\frac{1}{2}\right\}^\frac{1}{2}}.
\end{align}
In the limit $\tv \rightarrow \infty,$ one obtains the longitudinal correlation
function 
\begin{align}
\lim_{\tv \rightarrow \infty}\langle\phi(\txpar,0)\phi(\boldsymbol 0) \rangle
&\sim \frac{1}{4\sqrt{2} \pi }\int
d\tqpar \frac{e^{i \tqpar \txpar}}{(1+4\tv^2\tqpar^2)^\frac{3}{4}},\label{eqn:
}\\
&= \frac{-2^{5/4}}{ \tv \pi\Gamma(-1/4)}\left(\frac{|\tilde x_\parallel|}{2\tilde
\vel}\right)^{1/4}K_{-1/4}\left(\frac{|\tilde x_\parallel|}{2\tilde \vel}\right),
\end{align}
which gives  (\ref{eqn: longitudinal_2d_corr}).

\bibliographystyle{apsrev}
\bibliography{EMT_PRE_March}
\end{document}